\documentclass[aps,showpacs,preprintnumbers,amsmath,amssymb,floatfix]{revtex4}

\usepackage[mathscr]{eucal}
\usepackage{amssymb}
\usepackage{amsmath}
\usepackage{amscd}
\usepackage{bm}
\usepackage[dvips]{graphicx}

\usepackage{color}

\def\pderiv#1#2{\frac{\partial#1}{\partial#2}}
\def\pderivop#1{\frac{\partial}{\partial#1}}
\def\pnderivop#1#2{\frac{\partial^{#2}}{\partial#1^{#2}}}

\def\pnderivM#1#2#3{\frac{\partial^2 #1}{\partial #2 \partial #3}}
\def\d3V#1#2{\frac{d^{#1}#2}{(2\pi)^{#1}}}
\def\HALF{\frac{1}{2}}

\def\Xa{\bm{X}^{(\alpha)}}

\def\Xat#1{\bm{X}^{(\alpha)}(#1)}

\def\dotXat#1{\dot{\bm{X}}^{(\alpha)}(#1)}

\begin{document}

\title{Phase coherence in an ensemble of uncoupled limit-cycle
  oscillators receiving common Poisson impulses}
\author{Kensuke Arai$^1$\footnote{E-mail: arai@ton.scphys.kyoto-u.ac.jp\\
    URL: http://www.ton.scphys.kyoto-u.ac.jp/$\tilde{\ }$nonlinear}
 and Hiroya Nakao$^{1,2}$ }
\affiliation{Department of Physics, Kyoto University, Kyoto 606-8502,
  Japan$^1$ \\
  Abteilung Physikalische Chemie, Fritz-Haber-Institut der Max-Planck-Gesellschaft, Faradayweg 4-6,
  14195 Berlin, Germany$^2$
}
\date{\today}
             
\begin{abstract}
  An ensemble of uncoupled limit-cycle oscillators receiving common
  Poisson impulses shows a range of non-trivial behavior, from
  synchronization, desynchronization, to clustering.
  The group behavior that arises in the ensemble can be predicted from
  the phase response of a single oscillator to a given
  impulsive perturbation.
  We present a theory based on phase reduction of a jump stochastic
  process describing a Poisson-driven limit-cycle oscillator, and
  verify the results through numerical simulations and electric
  circuit experiments.
  We also give a geometrical interpretation of the synchronizing
  mechanism, a perturbative expansion to the stationary phase
  distribution, and the diffusion limit of our jump stochastic model.
\end{abstract}


\maketitle
 
\section{Introduction}
The improvement in response reproducibility of a system receiving
identical fluctuating drive has received much attention recently
~\cite{Roy,Mainen-Sejnowski,Royama,Binder-Powers,Galan,Toral,Zhou-Kurths,
  Pakdaman,Teramae-Dan,Goldobin-Pikovsky,Nagai-Nakao,Nakao-Arai,
  Nakao-Arai-Kawamura,Lin, Pikovsky-R-K1}.
With a constant input signal, many systems composed of identical
elements in an ensemble show unreliable response due to noise after
starting from similar initial conditions, while a fluctuating drive
greatly improves response reproducibility.  This general phenomenon is
observed in various guises throughout nature.  Uncoupled lasers driven
by a common master laser with a fluctuating output produce highly
correlated output intensities~\cite{Roy}.  In the rat neocortical
neuron~\cite{Mainen-Sejnowski}, action potential (spike) generation
times over many trials coincide when a fluctuating current is
delivered to the soma.  Synchronized firings \begin{it}in vivo\end{it}
in the cat spinal motoneurons~\cite{Binder-Powers}, and in neurons of
the olfactory bulbs in mice~\cite{Galan} have also been reported.
Population density correlation among spatially separated species,
known as the Moran effect~\cite{Royama}, is a general phenomenon seen
in many different organisms in the troposphere.  The phenomenon has
also been found in chaotic oscillators~\cite{Toral,Zhou-Kurths}, where
synchronization of the generalized phase has been observed, and is
known as noise-induced chaotic synchronization.
Common forcing may also lead to a decrease in response
  reproducibility~\cite{Pikovsky-R-K1,Goldobin-Pikovsky,Nagai-Nakao,Nakao-Arai,Lin,Tass,Hudson}.

\begin{figure}[htbp]
  \centering
  \includegraphics*[width=0.85\hsize]{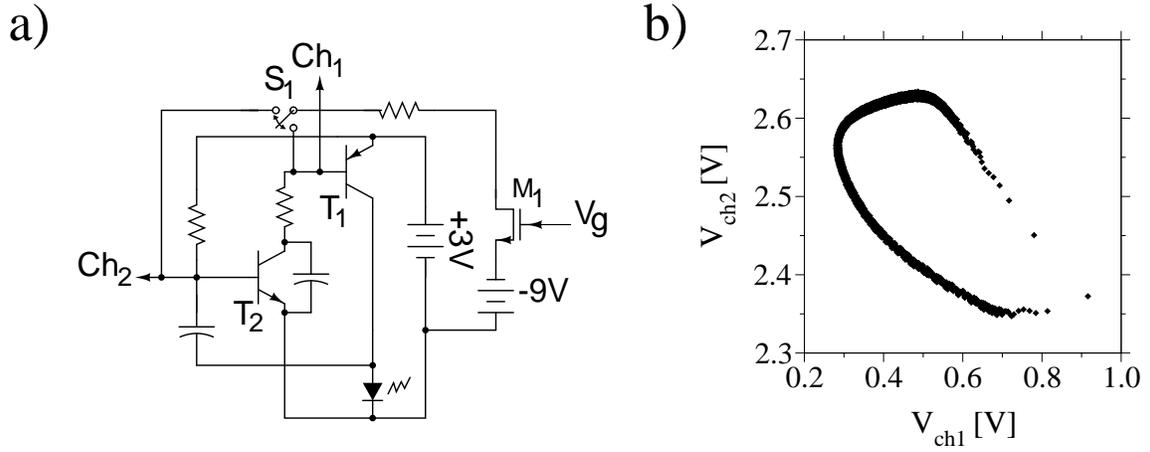}
  \caption{(a) Diagram of electrical circuit with limit-cycle
    behavior.  Computer-generated impulses control $V_g$, which turns
    MOSFET $\mbox{M}_1$ current source on/off.  Switch $\mbox{S}_1$
    allows us to send common impulse to either $Ch_1$ or $Ch_2$.  (b)
    Limit cycle of electrical circuit produce by measuring voltages at
    $Ch_1$ or $Ch_2$ as given in (a). }
  \label{Fig:ExptLC}
\end{figure}

Let us first illustrate with examples the phase coherence phenomenon
induced by fluctuating drive.
Figures~\ref{Fig:FHNLC} and~\ref{Fig:ExptLC} show the orbit of
unperturbed limit-cycle oscillators obtained numerically and
experimentally, and Figs.~\ref{Fig:FHNRasterPlots}
and~\ref{Fig:ExptRasterPlots} show the effect of common impulses on an
ensemble of such oscillators.
Though the oscillators do not interact with each other, they
exhibit phase synchronization and desynchronization
\footnote{This ``desynchronization'' does not mean merely passive
  phase diffusion due to noises.  It means active desynchronization
  due to the impulse-induced orbital instability, which may also be
  called stochastic chaos~\cite{Pakdaman}.}
depending on the impulse intensity and oscillator characteristics.
Knowledge of the response of the oscillator to an impulsive
perturbation is sufficient to understand the observed coherence
phenomena, as we will clarify in this paper.

Several previous studies have demonstrated this phenomenon for various
driving signals and
oscillators~\cite{Teramae-Dan,Goldobin-Pikovsky,Nagai-Nakao,Nakao-Arai,
  Nakao-Arai-Kawamura,Lin} by using the phase reduction method for
limit cycles~\cite{Winfree,Kuramoto}.
In particular, the case where the fluctuating signal is a
  sequence of random impulses has been investigated in Pikovsky,
  Rosenblum and Kurths (\cite{Pikovsky-R-K1}, Sec. 15 and references
  therein) and by ourselves~\cite{Nakao-Arai}.
Using random phase maps, it is argued that synchronization
  always occurs for general limit cycles when sufficiently weak
  additive random impulses are given, and desynchronization can also
  occur when the impulse intensity is finite.

In this article, we generalize our previous argument through a refined
formulation in terms of Poisson-driven Markov processes~\cite{Snyder},
also known as jump processes~\cite{Hanson}, which enables us to
systematically perform phase reduction and linear stability analysis
of our impulse-driven oscillators for general multiplicative coupling.
The oscillator response to the impulses is expressed as a function
called the phase response curve (PRC)~\cite{Winfree,Kuramoto},
  which is a very basic quantity of limit-cycle oscillators measured
  in every experiment, and the coherent behavior that arises from
the common impulse can be deduced once the PRC is obtained.
We present a criteria for predicting when each of the above coherence
phenomenon can be expected from the application of common impulses,
and test the predictions quantitatively using numerical simulation and
an electric circuit experiment.
We measure the PRC for a typical limit-cycle oscillator described by a
set of ordinary differential equations, and for an electrical
limit-cycle oscillator.
We then compare the rate of synchronization or desynchronization as
predicted by the Lyapunov exponent obtained from the PRC in both
numerical simulation and experiment.
We also give a simple geometrical explanation of the synchronization
as a consequence of the stability of the limit cycle, a perturbative
expansion to the phase distribution of a single oscillator, and a
derivation of the diffusion limit of the jump process describing our
impulse-driven limit-cycle oscillators in the Appendix.

\section{Theory}

In this section, we present linear stability analysis of the
impulse-driven oscillator through phase reduction of the jump
stochastic differential equation describing the model.
Similar analyses have been performed based on random phase map
  description for a simple sinusoidal phase map in
  Ref.~\cite{Pikovsky-R-K1}, Sec. 15, and for more general phase maps
  derived from the phase reduction of limit cycles in
  Ref.~\cite{Nakao-Arai}, which gave formulas relating the Lyapunov
  exponent to the functional form of the phase maps and predicted
  synchronization and desynchronization of uncoupled oscillators
  driven by common random additive impulses.
Our reformulation based on the stochastic jump process given in
  this section provides a simple and systematic treatment of the
  impulse-driven limit cycle oscillator for general multiplicative
  coupling, which quantitatively relates the Lyapunov exponents, the
  PRCs, and the phase-space structure (isochrons~\cite{Winfree}) in a
  mathematically transparent way, starting from a general dynamical
  equation describing impulse-driven limit cycles.  It also
  incorporates the effect of different stochastic interpretations for
  the random impulses.  Using our new formulation, we derive the
  classical results in a more general way and argue the possibility of
  clustered states for multiplicative impulses.

\subsection{Model}

We consider an $N$-oscillator ensemble receiving a common sequence of
random Poisson impulses.  The equation for the $\alpha$-th oscillator
is
\begin{align}
  \label{OriginalODE}
  \dotXat{t} = \bm{F}(\Xa) +
  \sum_{n = 1}^{N(t)} \bm{\sigma}(\Xa, \bm{c}_n) h(t - t_n),
\end{align}
where $\alpha = 1, \cdots, N$, $\Xat{t} \in \bm{R}^M$ is the
oscillator state at time $t$, $\bm{F} (\Xa):\bm{R}^M \to \bm{R}^M$ the
dynamics of a single oscillator, $N(t)$ the number of received
impulses up to time $t$, ${t_n}$ the arrival time of the $n$-th
impulse, $\bm{c}_n \in \bm{R}^K$ the intensity and direction, or {\em
  mark}~\cite{Snyder,Hanson}, of the $n$-th impulse, $\bm{\sigma}(\Xa,
\bm{c}):\bm{R}^M \times \bm{R}^K \to \bm{R}^M$ is the coupling
function describing the effect of an impulse to $\Xat{t}$, and $h(t -
t_n)$ is the unit impulse ($\int^{\infty}_{-\infty} h(t - t_n) dt =
1$) whose waveform is localized at the event time $t_n$.
We denote the rate of the Poisson impulses as $\lambda$, namely, the
mean interval between the impulses is $1 / \lambda$.

In the absence of impulses, we assume that each oscillator obeys the
same dynamics, with a single stable limit cycle solution $\bm{X}_0(t)$
of period $T$ in phase space.  In the following, we omit the
oscillator index $\alpha$ as the ensemble is composed of identical
uncoupled elements, and our discussion involves only the linear
stability of an individual oscillator.

\subsection{Jump stochastic differential equation}

We pose the problem as a Poisson-driven Markov process, or {\em
    stochastic jump process.}  We regard the impulse as an event of
zero-temporal width, so that the temporal correlation of the 
  impulses vanishes.  The resulting discontinuous oscillator dynamics
can be described by a stochastic-integro differential equation,
  or {\em jump stochastic differential equation (jump SDE)}, for a
marked Poisson point process~\cite{Snyder,Hanson}.  Some
properties of the jump SDE are given in Appendix A.

There are two ways we may interpret the impulse term in the original
ordinary differential equation (\ref{OriginalODE}), which we call Ito
and Stratonovich pictures for convenience.  The Ito-picture assumes
the impulsive term in the original Eq.~(\ref{OriginalODE}) as being
point-like impulses, which gives rise to discontinuous system
  dynamics with jumps.
The Stratonovich-picture assumes the impulsive term in the original
Eq.~(\ref{OriginalODE}) as a limit of short but continuous
  waveforms of non-zero width, thus requiring that we find the limit
of the system response as the impulses are shrunk to $0$ width.

Both pictures lead to the same jump SDE for the phase variable:
\begin{equation}
  \label{GeneralSIDE}
  d\bm{X}(t) = \bm{F}(\bm{X})dt + \int_{\bm{c}} \bm{g}(\bm{X}, \bm{c}) M(dt, d\bm{c}),
\end{equation}
which is always interpreted in the Ito sense. Here, $M(dt,
d\bm{c})$ represents a Poisson random measure, which gives the number
of incident points during $[t, t+dt]$ having the mark $[\bm{c},
\bm{c}+d\bm{c}]$, whose expectation is
\begin{equation}
  \langle M(dt, d\bm{c}) \rangle = \lambda dt p(\bm{c}) d\bm{c},  
\end{equation}
where $\lambda$ is the rate of the Poisson process and
  $p(\bm{c})$ is the probability density function (PDF) of the marks
  $\bm{c}$, and the integral is over the mark
  space~(Ref.~\cite{Snyder,Hanson}, Appendix A).
The oscillator response $\bm{g}(\bm{X}, \bm{c})$ to a given
  impulse with mark $\bm{c}$ is different between the two pictures
  when the effect of the impulse $\bm{\sigma}(\bm{X}, \bm{c})$ is
  multiplicative.
When the full oscillator dynamics is reduced to phase dynamics,
  the phase response is slightly altered, as we
  see later.

\subsubsection{Ito picture}

We interpret the common impulse $h(t - t_n)$ in the original
Eq.~(\ref{OriginalODE}) to be zero-width from the outset.  When an impulse $\bm{c}$ is received at time $t$, the state of the
oscillator jumps discontinuously from $\bm{X}$ to $\bm{X} +
\bm{\sigma}(\bm{X}, \bm{c})$.  This interpretation gives a jump SDE of the form
\begin{equation}
  \label{Ito}
  d\bm{X}(t) = \bm{F}(\bm{X})dt + \int_{\bm{c}} \bm{\sigma}(\bm{X}, \bm{c}) M(dt, d\bm{c}).
\end{equation}
Thus, $\bm{g}(\bm{X}, \bm{c}) = \bm{\sigma}(\bm{X}, \bm{c})$.

\subsubsection{Stratonovich picture}
We interpret the impulse $h(t - t_n)$ in the original
Eq.~(\ref{OriginalODE}) as having a continuous waveform of
finite width and then shrink the impulse width to 0.
The resulting continuous process can be approximated by a 
  discontinuous jump SDE with a jump amplitude as determined
by the canonical form of the Wong-Zakai theorem for jump processes as shown by Marcus~\cite{Wong-Zakai,Marcus}.  Defining the differential operator
\begin{equation}
  \hat{D} = \sum_{l = 1}^{M} \sigma_l(\bm{X}, \bm{c}) \pderivop{X_l},
\end{equation}
the jump SDE is given by
\begin{equation}
  \label{WhiteNoiseLimit}
  d\bm{X}(t) =  \bm{F}(\bm{X}) dt + \int_{\bm{c}} \left[e^{\hat{D}}\bm{X} - \bm{X} \right] M(dt, d\bm{c}).
\end{equation}
Thus, $\bm{g}(\bm{X}, \bm{c}) = e^{\hat{D}}\bm{X} - \bm{X} $ in this
case.

Note that for additive impulses, $\bm{\sigma}(\bm{X}, \bm{c}) \equiv
\bm{\sigma}(\bm{c})$, Eq.~(\ref{Ito}) and Eq.~(\ref{WhiteNoiseLimit})
take the same form because $e^{\hat{D}}\bm{X} - \bm{X} =
\bm{\sigma}(\bm{c})$.  Therefore, the difference in stochastic
interpretation is not important in this case.
For the linear multiplicative case, $\sigma_k(\bm{X}, \bm{c}) = c_k
X_k$.  It is easy to check that $g_k(\bm{X}, \bm{c}) = (e^{c_k} -
1)X_k$.  This expression for the instantaneous jump magnitude
approximates the effect of an impulsive perturbation due to a short
but finite-width impulse whose intensity in each direction is $c_k$.
For more general multiplicative coupling, explicit expressions
  for $\bm{g}(\bm{X}, \bm{c})$ are difficult to obtain.

\subsection{Phase reduction}

To facilitate theoretical analysis, we apply phase
reduction~\cite{Winfree,Kuramoto} to Eq.~(\ref{GeneralSIDE}),
assuming that the average time-interval between jump events is long
compared to the relaxation time of the perturbation to the limit cycle
orbit.
An unperturbed oscillator executes periodic motion along its limit
cycle, so its state can be described by one phase variable $\phi(t) =
\phi(\bm{X}_0(t)) \in [0, 1)$ which constantly increases with
  frequency $\omega = 1 / T$, instead of the original $M$ variables.

Because the limit cycle is assumed to be globally stable, the orbit of
any initial point $P$ off of the limit cycle will
asymptotically approach the limit cycle.
Thus, we can extend the definition of the phase $\phi$ to the
  whole phase space except at phase singular sets by identifying the
set of points that asymptotically converge to the same orbit on the
limit cycle with the same phase, called the {\em
    isochron}~\cite{Winfree,Kuramoto}.
In practice, if the orbit approaches a point on the limit cycle with
phase $\phi$ to within some arbitrarily small distance after time
$\tau$, we define the asymptotic phase of the initial point $P$ to be
$(\phi - \tau / T)\mbox{ mod } 1$, where $T$ is the period of the
oscillator.

Now we perform phase reduction, which is mathematically a
change of variables describing the system from $\bm{X}$ to
$\phi = \phi(\bm{X})$, and is also an approximation of the function of
  $\bm{X}$ by the corresponding function of $\bm{X}_0(\phi)$.
In this transformation, every value of $\bm{X}$ in the neighborhood of
the limit-cycle attractor maps to a value of $\phi$ except at phase
singular sets.
Using the stochastic chain rule for the jump process 
  (Refs.~\cite{Snyder,Hanson}, Appendix A), we obtain
\begin{eqnarray}
  \label{NotClosedPhaseEQ}
  d\phi(t) = \omega dt + \int_{\bm{c}}
  \left[\phi\left(\bm{X} + \bm{g}(\bm{X}, \bm{c})\right) - \phi(\bm{X}) \right] M(dt, d\bm{c}),
\end{eqnarray}
which is not yet a closed equation for $\phi$.
The map $\bm{X} \to \bm{X} + \bm{g}(\bm{X}, \bm{c})$ describes the
effect of an impulse received when an oscillator is at $\bm{X}$.
Since we assume that the average interval between impulses is longer than
the relaxation time of the perturbation to the limit cycle, we can
evaluate the function of $\bm{X}$ using values of $\bm{X}_0$ on the
limit cycle, for which the mapping $\phi \to \bm{X}_0$ is well
defined.  Replacing the $\bm{X}$ with $\bm{X}_0(\phi)$, we obtain
\begin{eqnarray}
  \label{PhaseSIDE}
  d\phi(t) & \cong & \omega dt +
  \int_{\bm{c}} \left[\phi\left(\bm{X}_0(\phi) + \bm{g}(\bm{X}_0(\phi), \bm{c})\right) - \phi \right]
  M(dt, d\bm{c}) \cr \cr
  & = & \omega dt + \int_{\bm{c}} G(\phi, \bm{c}) M(dt, d\bm{c}),
\end{eqnarray}
which is now a closed equation for $\phi$.
Here we introduced a function
\begin{equation}
  G(\phi, \bm{c})  
  = \phi\left(\bm{X}_0(\phi) + \bm{g}(\bm{X}_0(\phi), \bm{c})\right) - \phi,
\end{equation}
which is the phase response curve (PRC) representing the change in
asymptotic phase relative to an unperturbed oscillator caused by an
impulse with mark $\bm{c}$ received at phase $\phi$.
Since we consider a continuous dynamical system, $G(\phi,
\bm{c})$ is continuous and periodic in $\phi$.
If the jump amplitude $\bm{g}(\bm{X}_0(\phi), \bm{c})$ is small,
$G(\phi, \bm{c})$ can be approximated as
\begin{equation}
  G(\phi, \bm{c}) \cong \bm{Z}(\phi) \cdot \bm{g}(\bm{X}_0(\phi), \bm{c}),
\end{equation}
where
\begin{equation}
  \bm{Z}(\phi) = \mbox{grad}_{\bm {X}} \phi({\bm X}) |_{\bm{X} = \bm{X}_0(\phi)}
  \label{PhaseSensitivity}
\end{equation}
is the well-known phase sensitivity function that represents linear
sensitivity of the phase to infinitesimal
perturbations~\cite{Winfree,Kuramoto}.
In general, this linear relationship holds only for very weak
  impulses (see e.g. \cite{Nakao-Arai}).  The PRC easily becomes
  a largely fluctuating function, which can even become jagged,
  e.g. near the bifurcation point.

Note that for our phase-reduction analysis of Poisson-driven
  limit cycles, the impulse intensity $\bm{c}$ need not be
  infinitesimally weak provided that the inter impulse intervals are
  sufficiently long, in contrast to the conventional phase-reduction
  analysis of limit cycles driven by continuous
  signals~\cite{Teramae-Dan,Goldobin-Pikovsky,Nakao-Arai-Kawamura}.
  It holds for largely fluctuating PRCs as well.  By virtue of this
  fact, we can analyze desynchronization effect of random signals
  within the framework of a one-dimensional phase model, in contrast to
  Ref.~\cite{Goldobin-Pikovsky}, as we discuss below.


\subsection{Linear stability of the synchronized state}

Whether synchronization occurs among the oscillators depends on the
stability of the synchronized state.
To investigate the linear stability of the synchronized state with the
application of common impulses, we focus on the time evolution of a
small phase difference $\psi = \tilde{\phi} - \phi$ between phase
$\phi$ and $\tilde{\phi}$ of two nearby orbits.  From
Eq.~(\ref{PhaseSIDE}), the linearized evolution equation for $\psi$ is
given by
\begin{equation}\label{PhaseDiffEq}
  d\psi(t) = \int_{\bm{c}} \psi \frac{\partial}{\partial \phi} G(\phi, \bm{c}) M(dt, d\bm{c}).
\end{equation}
Now we change variables to the logarithm of the absolute value
of the phase difference, $y = \mbox{log}| \psi|$.  Using the
stochastic chain rule for the jump process, we obtain
\begin{eqnarray}
  dy(t) & = & \int_{\bm{c}} \left[ \mbox{log} | \psi + \psi G'(\phi, \bm{c}) | - \mbox{log} |\psi | \right] M(dt, d\bm{c})
  \cr \cr
  &= & \int_{\bm{c}} \mbox{log}  | 1 + G'(\phi, \bm{c}) | M(dt, d\bm{c}),
\end{eqnarray}
where the prime($'$) denotes partial derivative by $\phi$.  The average growth rate of the small phase difference is characterized
by the Lyapunov exponent $\Lambda$.
In this case, it is defined as
\begin{eqnarray}\label{LongTimeAvg}
  \Lambda &=&
  \lim_{T \to \infty} 
  \frac{1}{T} \log \left|\frac{\psi(T)}{\psi(0)} \right|
  =
  \lim_{T \to \infty} \frac{y(T) - y(0)}{T} 
  =
  \lim_{T \to \infty} \frac{1}{T} \int_0^T dy(t).
\end{eqnarray}
When $\Lambda$ is negative, the initial phase difference decays
exponentially, so that the synchronized state is linearly stable.  As usual, we postulate that the growing and shrinking of the
phase difference is ergodic, namely, the long-time average slope of
  $y(t)$ coincides with the ensemble average of its local slope,
\begin{eqnarray}
  \lim_{T \to \infty} \frac{1}{T} \int_0^T dy(t) = \frac{\langle dy(t) \rangle}{dt},
\end{eqnarray}
where $\langle \cdots \rangle$ denotes ensemble average over the
  marked Poisson process.  Note that an individual increment $dy(t)$
  may exhibit discontinuous jumps of $O(1)$, but the ensemble average
  $\langle dy(t) \rangle$ is always of $O(dt)$.
The expectation of the right-hand-side is calculated by
  replacing the dynamics of $\phi$ with the single-oscillator
  stationary PDF $p(\phi)$ of $\phi$, as

\begin{eqnarray}\label{Ensembledy}
  \langle dy(t) \rangle
  &=&
  \left\langle \int_{\bm{c}} \log\left| 1 + G'(\phi(t), \bm{c})\right| M(dt, d\bm{c}) \right\rangle \cr \cr
  &=&
  \int d\phi p(\phi) \int_{\bm{c}} \log\left| 1 + G'(\phi, \bm{c})\right| \langle M(dt, d\bm{c}) \rangle \cr \cr
  &=&
  \lambda dt \int_{0}^{1} d\phi p(\phi) \int_{\bm{c}} d\bm{c} p(\bm{c}) \log\left| 1 + G'(\phi, \bm{c})\right|, \;\;\;\;\;\;\;\;
\end{eqnarray}
where in the second line, the ensemble average is separated into a conditional expectation with fixed $\phi$ and average over $p(\phi)$ because the integration variable and stochastic driving term are statistically independent.
Thus, the Lyapunov exponent $\Lambda$ is obtained as
\begin{eqnarray}
  \label{LyapunovFromGFull}
  \Lambda
  & = & \lambda \int_{0}^{1} d\phi p(\phi) \int_{\bm{c}} d\bm{c} p(\bm{c})
  \log\left| 1 + G'(\phi, \bm{c})\right|,
\end{eqnarray} 
which generalizes the result obtained by random phase maps in
  Refs.~\cite{Pikovsky-R-K1} and ~\cite{Nakao-Arai} for general
  multiplicative coupling.
The sign of the Lyapunov exponent depends on the shape of the PRC,
$G(\phi, \bm{c})$.
When $G'(\phi, \bm{c}) < -2$ or $G'(\phi, \bm{c}) > 0$, the integrand,
which gives the instantaneous growth rate of $\psi(t)$ at
  $\phi$, is positive.  Such regions tend to expand the phase
difference between two orbits.
When $-2 < G'(\phi, \bm{c}) < 0$, the integrand is negative, and the
phase difference between two orbits tends to shrink.
Linear stability of the synchronized state is determined by the
overall balance between these two effects.

For weak impulses, we can further simplify
Eq.~(\ref{LyapunovFromGFull}) by assuming that, on average, an
oscillator is equally distributed on the limit-cycle, so $p(\phi) =
1$.
This is a reasonable assumption in most cases where the effect
  of impulses are small~\cite{Nakao-Arai}, as we discuss later in
Appendix B.
Under this approximation, the Lyapunov exponent $\Lambda$ can be
simplified as
\begin{equation}
  \label{LyapunovFromG}
  \Lambda = \lambda \int_{0}^{1} d\phi
  \int_{\bm{c}} d\bm{c} p(\bm{c}) \mbox{log}  | 1 + G'(\phi, \bm{c}) |.
\end{equation}
Now, if the impulses are weak and the variation of the PRC $G(\phi,
\bm{c})$ is sufficiently small in such a way that $-1 < G'(\phi,
\bm{c})$ is always satisfied for all $\phi$ and $\bm{c}$, i.e. when
$\phi + G(\phi, \bm{c})$ is a monotonically increasing function,
$\Lambda$ can be bounded from above as~\cite{Nagai-Nakao,Nakao-Arai}
\begin{eqnarray}
  \Lambda
  &\leq&
  \lambda \int_{0}^{1} d\phi \int_{\bm{c}} d\bm{c} G'(\phi, \bm{c}) p(\bm{c}) 
  =
  \lambda \int_{\bm{c}} p(\bm{c}) d\bm{c}  
  \left[ G(\phi, \bm{c}) \right]_{\phi = 0}^{\phi = 1}
  =
  0,
\end{eqnarray}
where we utilized the periodicity of $G(\phi, \bm{c})$ in $\phi$ and
the inequality $\log(1+x) \leq x$.
Thus, for weak impulses, small perturbations are always statistically
stabilized when averaged over the limit cycle, so that common impulses
shrink the phase difference, irrespective of the details of the
oscillator.  A geometrical interpretation of this stabilization
mechanism is given in Appendix C.

By a Taylor expansion of Eq.~(\ref{LyapunovFromG}), the Lyapunov
exponent can be approximated for weak impulse as
\begin{eqnarray}
  \label{LyapunovFromG_TaylorExp}
  \Lambda
  \cong
  \lambda \int_0^1 d\phi \int_{\bm{c}} d\bm{c} p(\bm{c})
  \left[ G'(\phi, \bm{c}) - \frac{G'(\phi, \bm{c})^2}{2} \right]
  =
  - \frac{\lambda}{2} \int_0^1 d\phi \int_{\bm{c}} d\bm{c} p(\bm{c}) G'(\phi, \bm{c})^2
  \leq 
  0,
\end{eqnarray}
where the first term drops out due to the periodicity of $G(\phi,
\bm{c})$.  At the lowest order approximation, $G(\phi, \bm{c})
  \simeq {\bm Z}(\phi) \cdot \bm{g}(\phi, \bm{c}) \simeq
  \sum_{k=1}^{M} Z_k(\phi) \sigma_k(\phi, \bm{c})$, the approximate
Lyapunov exponent $\Lambda$ is obtained as
\begin{eqnarray}
  \Lambda
  &=& -\frac{\lambda}{2} \sum_{k=1}^{M} \sum_{l=1}^{M}
  \int_0^1 d\phi 
  [
  Z_k'(\phi) Z_l'(\phi) \langle \sigma_k \sigma_l \rangle_{\bm{c}}(\phi) \cr \cr
  &&+
  2 Z_k'(\phi) Z_l(\phi) \langle \sigma_k \sigma_l' \rangle_{\bm{c}}(\phi)
  +
  Z_k(\phi) Z_l(\phi) \langle \sigma_k' \sigma_l' \rangle_{\bm{c}}(\phi)
  ]
\end{eqnarray}
for both Ito and Stratonovich pictures of Eq.~(\ref{OriginalODE}),
where we introduced correlation functions $\langle \sigma_k \sigma_l
\rangle_{\bm{c}}(\phi) = \int_{\bm{c}} d\bm{c} p(\bm{c})
\sigma_k(\phi, \bm{c}) \sigma_l(\phi, \bm{c})$, etc.
As we derive in Appendix D, we obtain the same Lyapunov exponent in
the diffusion limit of the jump process.


\subsection{Clustered states}

If the PRC possesses a symmetry
\begin{eqnarray}
  G(\phi, \bm{c}) = G(\phi + \frac{1}{m}, \bm{c})
\end{eqnarray}
where $m \in \bm{N}$ and $1/m < 1$ (i.e. $m = 2, 3, 4,
  \cdots$), the stability of the synchronized state (zero phase
difference) also implies the existence of stable states separated in
phase by $1/m$.
Defining $\psi$ as a small deviation from such a state, we set
\begin{eqnarray}
  \psi = \phi' - \phi - \frac{1}{m}.
\end{eqnarray}
The linear stability analysis of a state separated in phase by $1/m$
yields the same Eq.~(\ref{PhaseDiffEq}) when utilizing PRC symmetry,
so that the resulting Lyapunov exponent $\Lambda$,
Eq.~(\ref{LyapunovFromG}), also takes the same value as that of the
synchronized state.

If $\Lambda$ is negative, the phase difference between two orbits can
stably take $m$ different values.
When many identical oscillators are driven by common impulses and also
by small independent disturbances in such a situation, the oscillators
will eventually split into $m$ clusters.
Any two oscillators inside the same group are synchronized, and those
belonging to different clusters will take one of $m-1$ phase
differences, $n/m$ where $n=1, \cdots, m-1$.
We call this an $m$-cluster state.
As we demonstrate later, symmetry in the original limit cycle results
in symmetry of the corresponding PRC, leading to cluster states.

\section{Numerical simulation}

In this section, we demonstrate synchronization, desynchronization,
and clustering induced by common impulses by numerical simulations,
and test our theoretical predictions quantitatively, using the
  FitzHugh-Nagumo (FHN) neural oscillator as an example.
The effect of common random signals for uncoupled FHN oscillators
  have been discussed for Gaussian signals in
  Ref.~\cite{Goldobin-Pikovsky} and for a random telegraphic signal in
  Ref.~\cite{Nagai-Nakao}.
However, quantitative comparison of the Lyapunov exponent
  predicted from the PRC with that directly measured from numerical
  simulations have not been fully done, especially in the
  desynchronization regime.
In this section, we measure the PRC and the separation rate of
  nearby trajectories directly by numerical simulations, and
  quantitatively confirm the theoretical prediction based on the
  one-dimensional phase model.
Results for the Stuart-Landau oscillator, which is qualitatively
  different from the FHN oscillator, is also given in Appendix E for
  comparison.

\subsection{Model}

For the simulation, we employ the FitzHugh-Nagumo (FHN) neural
oscillator~\cite{Koch} as an example, described by the following set
of equations:
\begin{eqnarray}
  \dot{u}(t) & = & \varepsilon (v  + a - b u), \cr 
  \dot{v}(t) & = & v - \frac{v^3}{3} - u + I_0 + \sigma(v, c) \sum_{n=1}^{N(t)} h(t - t_{n}) + \sqrt{D}\xi(t).
  \;\;\;\;\;\;\;\;\;
\end{eqnarray}
Here, parameters $\varepsilon, a, b$ are fixed at $\varepsilon
= 0.08$, $a = 0.7$, $b = 0.8$, and we use the parameter $I_0$
to control the oscillator characteristics.  The last two terms of the
equation for $v$ describe impulses and noises.
The $h(t)$ is a common impulse with unit intensity, generated by a Poisson point process of constant-rate $\lambda$.
The function $\sigma(v, c)$ describes $v$-dependent effect of the
impulse to the oscillator (for simplicity, we do not consider the case where $c$ takes
  multiple values in the following).  The $\xi(t)$ is a Gaussian white noise
with intensity $D$ describing independent disturbances to the
oscillators.

\begin{figure}[htbp]
  \centering
  \includegraphics*[width=0.8\hsize]{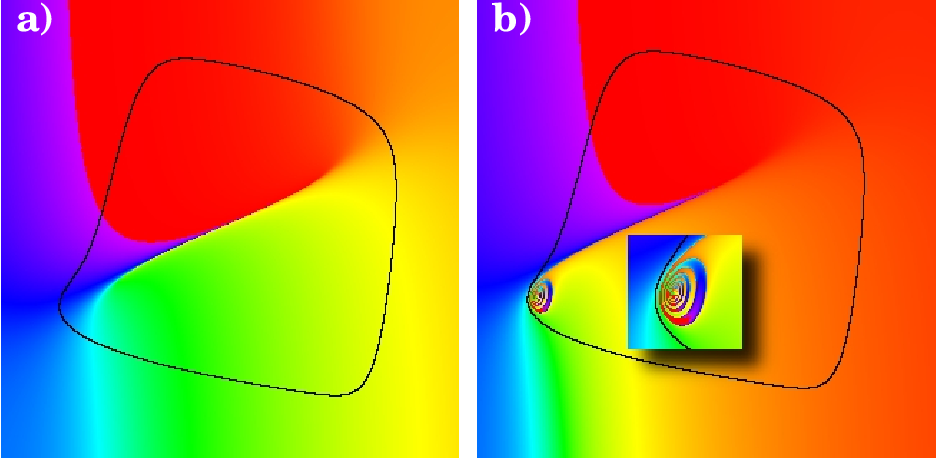}
  \caption{(Color online) Asymptotic phase for FHN oscillator with limit cycle in
    black.  $I_0 = 0.8$ and $0.34$ in (a), (b), respectively.  The
    center of the spiral in (b) occurs at the intersection of the
    nullclines, and is the remnant of a destabilized fixed point as
    the oscillator passes through a subcritical Hopf bifurcation where
    $I_0$ is the bifurcation parameter.}
  \label{Fig:FHNLC}
\end{figure}

When both external disturbances are zero, a limit cycle exists for $I_0 \in
[0.331, 1.419]$, which is created by a subcritical Hopf bifurcation at
either limits of $I_0$.
Figure~\ref{Fig:FHNLC} displays portraits of asymptotic phase for two
values of $I_0$ in the absence of impulses and noises.  At $I_0
= 0.8$, the period of the limit cycle is $T \approx 36.52$, and the
oscillator has a smooth phase portrait as shown in
Fig.~\ref{Fig:FHNLC}(a).
Near the bifurcation point, $I_0 = 0.34$, the period of the
  limit cycle is $T\approx 46.79$.  The remnants of the destabilized
fixed point exist at this parameter, as seen in
Fig.~\ref{Fig:FHNLC}(b).
We define the origin of phase $\phi$ as the point where the variable
$v$ passes through the $v=0.9$ line from below, where the FHN
oscillator appears to emit a neuronal action potential.

\subsection{Phase response curves}

\begin{figure}[htbp]
  \centering
  \includegraphics*[width=0.8\hsize]{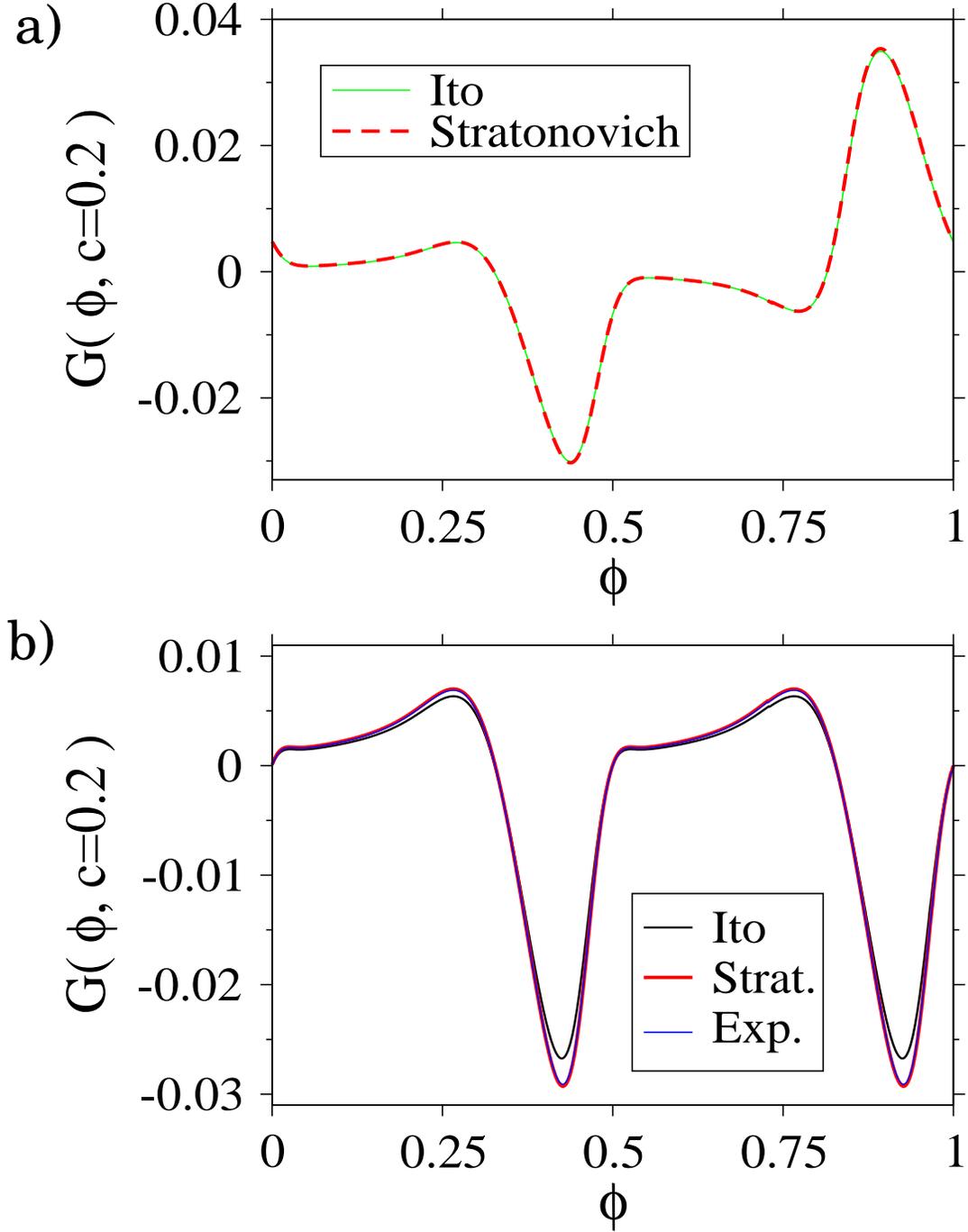}
  \caption{(Color online) 
    Comparison of Ito vs. Stratonovich interpretations of Eq.~(\ref{OriginalODE})
    on the PRC $G(\phi, c)$ of FHN for an impulse whose jump size is
    $c=0.2$.
    (a) Additive impulse ($\sigma(v, c) = c$) and
    (b) Linear multiplicative impulse ($\sigma(v, c) = cv$).
    The curve ``Ito'' is calculated by affecting a discontinuous jump, i.e.
    impulse duration is $0$.  The curve ``Stratonovich'' and ``Strat.'' is calculated by
    continuously changing $v$ using a narrow rectangular
    waveform of temporal width $0.0002$.
    The curve "Exp." is calculated using the Wang-Zakai-Marcus
    approximation for the continuous narrow impulse, namely,
    discontinuously changing the orbit by an amount $g(v, c) = (e^c
    - 1) v$.  }
  \label{Fig:ItoVsStratonovich}
\end{figure}

In simulation, we may use two algorithms corresponding to Ito and
Stratonovich pictures of Eq.~(\ref{OriginalODE}).
If we treat the impulses as point events, allowing discontinuous
  jumps of the orbit of the oscillator, the results correspond to the Ito
picture, Eq.~(\ref{Ito}).
If we directly integrate short but non-zero-width
  continuous impulses, the results correspond to
the Stratonovich picture, Eq.~(\ref{WhiteNoiseLimit}).  When the
effect of impulses are multiplicative, the PRC is different between
the two interpretations.
Figure~\ref{Fig:ItoVsStratonovich} displays the PRCs obtained using
Ito and Stratonovich pictures for additive impulses ($\sigma(v, c) = c$)
and for linear multiplicative impulses ($\sigma(v, c) = cv$).
For additive impulses, both algorithms give the same PRCs.
In contrast, for multiplicative impulses, it is seen that a difference
in the pictures has a small but visible effect on the PRC.  Of course,
the PRC obtained using Wong-Zakai-Marcus approximation coincides with
that obtained in the Stratonovich picture.
All the following simulations are done using the Stratonovich picture
of the original Eq.~(\ref{OriginalODE}), namely
Eq.~(\ref{WhiteNoiseLimit}), using the Wong-Zakai-Marcus approximation
of a continuous physical jump.

At $I_0 = 0.8$, the PRC for additive impulses is a smooth periodic
function as shown in Fig.~\ref{Fig:FHNPRC}(a) for all values of the
impulse intensity $c$, corresponding to the smooth phase portrait as
shown in Fig.~\ref{Fig:FHNLC}(a).
In contrast, at $I_0 = 0.34$, the PRC can become jagged as shown in
Fig.~\ref{Fig:FHNPRC}(b) for the impulse intensity $c$ in a certain
range.
This reflects the existence of the unstable focus, which looks
  like a spiral in Fig.~\ref{Fig:FHNLC}(b).
If the impulse intensity is in such a range that the orbit on the
limit cycle is kicked into this region, an initial phase difference
can grow quickly because the asymptotic phase in that region varies so
rapidly.

\subsection{Synchronization, desynchronization, and clustering}

Raster plots of $N=20$ uncoupled FHN oscillators,
Fig.~\ref{Fig:FHNRasterPlots}, which indicate times that an oscillator
passed through $\phi = 0$, offer a qualitative picture of the
phenomenon.  Whether the impulse is introduced additively or
multiplicatively, we find system and impulse parameters where phase
synchronization occurs and where it does not.
FHN oscillator has a large parameter range in $I_0$ and $c$ where
common impulses cause the oscillators to synchronize, as shown in
Fig.~\ref{Fig:FHNRasterPlots}(a).
However, near the bifurcation, common impulses sometimes accelerate
the desynchronization of the oscillators, as shown in
Fig.~\ref{Fig:FHNRasterPlots}(b).
These distinct behavior can be understood by examining the PRC of the
FHN oscillator at each parameter value.

\begin{figure}[htbp]
  \centering
  \includegraphics*[width=0.8\hsize]{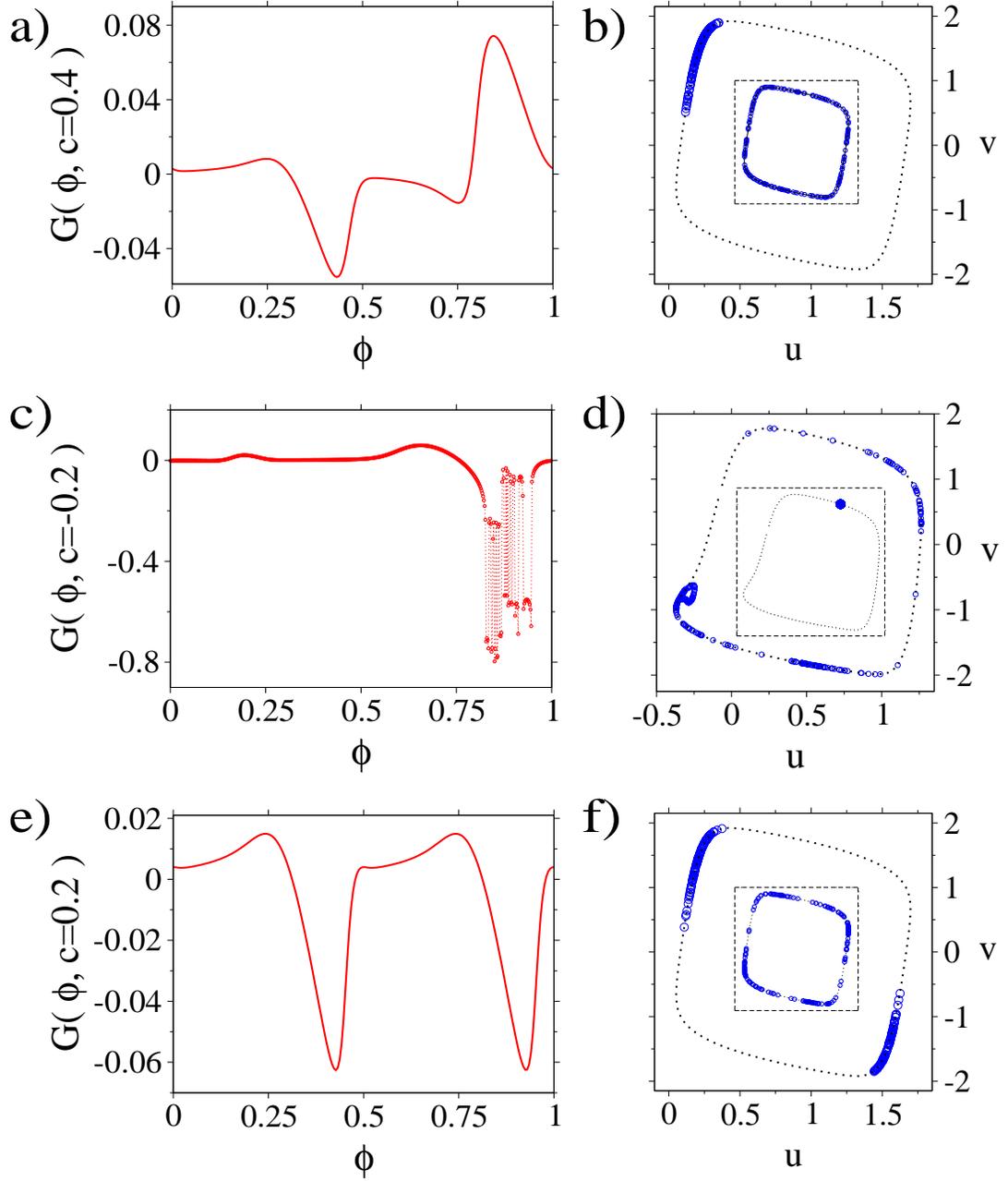}
  \caption{(Color online) 
    Various PRCs and coherent states of FHN oscillators.
    (a), (c), (e) show PRCs $G(\phi, c)$, and (b), (d), (f) show
    corresponding phase-space diagrams of 200 oscillators a
    sufficient time after the initial condition shown in the
    insets.
    Poisson rate is $\lambda = 1/4T$ for (a), (b), and (f), and
    independent noise is $D = 5 \times 10^{-6}$ for (b) and (f),
    and $D = 9 \times 10^{-9}$ for (d).
    The control parameter is $I_0 = 0.875$ for (a), (b), (e),
    (f), and $I_0 = 0.34$ for (c) and (d).
    For weak additive impulse, the PRC $G(\phi, c)$ is a
    periodic function, (a), and the 1-cluster (synchronized) state,
    (b), appears.
    At $I_0 = 0.34$, the PRC $G(\phi, c)$ becomes jagged when the
    additive impulse intensity is in a certain range as shown in (c),
    which often leads to common-impulse induced desynchronization, (d).
    For multiplicative impulse at $I_0 = 0.875$, a doubly periodic PRC
    $G(\phi, c)$, (e), leads to the 2-cluster state, (f).  }
  \label{Fig:GeneralCoherence}
\end{figure}

 Figure~\ref{Fig:GeneralCoherence} summarizes the relation
  between the shapes of the PRC and the dynamics of the oscillator
  ensemble in the phase space. 
For a smooth PRC with relatively small amplitudes obtained at
  $I_0 = 0.8$ for additive impulse ($\sigma(v, c) = c$) as shown in
  Fig.~\ref{Fig:GeneralCoherence}(a), the oscillators groups together
  on the limit cycle, namely, synchronize with each other, by the
  application of common impulses, Fig.~\ref{Fig:GeneralCoherence}(b).
For jagged PRC with strong amplitude fluctuations obtained by
  applying additive impulses near the bifurcation point $I_0 = 0.34$ as
  shown in Fig.~\ref{Fig:GeneralCoherence}(c), the oscillators undergo
  desynchronization by the application of common impulses and scatter
  along the limit cycle, as shown in
  Fig.~\ref{Fig:GeneralCoherence}(d).
At the midway point between the subcritical bifurcation near $I_0 =
0.875$, the limit cycle becomes symmetric about $v = 0$.
If the impulse is applied in a linear multiplicative way ($\sigma(v,
c) = cv$) in this $I_0$ region, the PRC becomes doubly periodic, as
shown in Fig.~\ref{Fig:GeneralCoherence}(e), so that $2$-cluster state
appears as shown in Fig.~\ref{Fig:GeneralCoherence}(f).
Even when the parameters of the oscillators are slightly
inhomogeneous, these dynamical behavior remain qualitatively unchanged
\footnote{Strictly speaking, synchronization between uncoupled
  oscillators due to common external drive is somewhat different from
  synchronization between coupled oscillators~\cite{Yoshimura}.
  In uncoupled cases, the long-time average frequency of each oscillator
  remains unchanged, so that the average frequency difference between
  two oscillators never vanishes.
  Therefore, the phase difference continues to increase, unlike
  coupled oscillators where the phase difference locks at a certain
  value.
  In uncoupled cases, the synchronization appears as the tendency
  for the phase difference to stay at a certain value between
  successive one-period slips of the phase difference.}.

In the present case for FHN oscillators, synchronization gradually
  occurs whereas desynchronization occurs suddenly due to the narrow
  jagged part of the PRC, typically obtained near the bifurcation
  point of the dynamics.
However, we emphasize that desynchronization is not limited to such
  pathological situations.  The PRC need not be rapidly fluctuating as long as
  intervals where the sufficiently steep slopes outweigh shallower
  slopes in Eq.~(\ref{LyapunovFromG}).
  As we see in the next section, our electrical oscillator is of this type, where desynchronization occurs even though the oscillator is far from a bifurcation and the PRC is smoothly.
  As an example of this from a well-known system, we include results from the Stuart-Landau
  oscillator in Appendix E.

In Ref.~\cite{Goldobin-Pikovsky}, the mechanism for
  desynchronization is analyzed for FHN oscillators near the
  bifurcation point driven by finite strength white Gaussian forcing
  utilizing a two-variable phase-amplitude model to take into account
  the non-trivial transverse deviation from the limit cycle.
In contrast, for our Poisson forcing case (and also for our
  previous case with random telegraphic forcing~\cite{Nagai-Nakao}),
  we can eliminate the relaxation dynamics of the amplitude and
  isolate the effects of the phase perturbation, because the
  time-scale of oscillator relaxation back to the limit-cycle is
  assumed to be much shorter than the average impulse interval.  We
  can thus stay within the framework of a single-variable phase model
  to describe both the synchronization and desynchronization, which
  elucidates the relation between the phase space structure
  (isochrons) and the desynchronization mechanism.

This desynchronization mechanism is qualitatively similar to the
  mechanism of the singular behavior in circadian clocks proposed by
  Winfree~\cite{Winfree2}, who argued that the attenuation of the
  circadian rhythm by an external stimulus is due to the
  desynchronization of multiple independent circadian oscillators
  being kicked into the unstable focus of the limit cycle.
  In Ref.~\cite{Tass}, application of such a desynchronization
  mechanism to neural populations are discussed.
  Based on the same idea, the sudden desynchronization seen in coupled
  chemical oscillators when a common perturbation is administered at
  the correct timing is also reported~\cite{Hudson}.
  Recently, Ukai {\it et al.}~\cite{Kobayashi} performed a very clear
  experiment of this phenomenon using genetically engineered
  photosensitive cells exhibiting circadian oscillations.

\subsection{Lyapunov exponents}

\begin{figure}[htbp]
  \centering
  \includegraphics*[width=0.8\hsize]{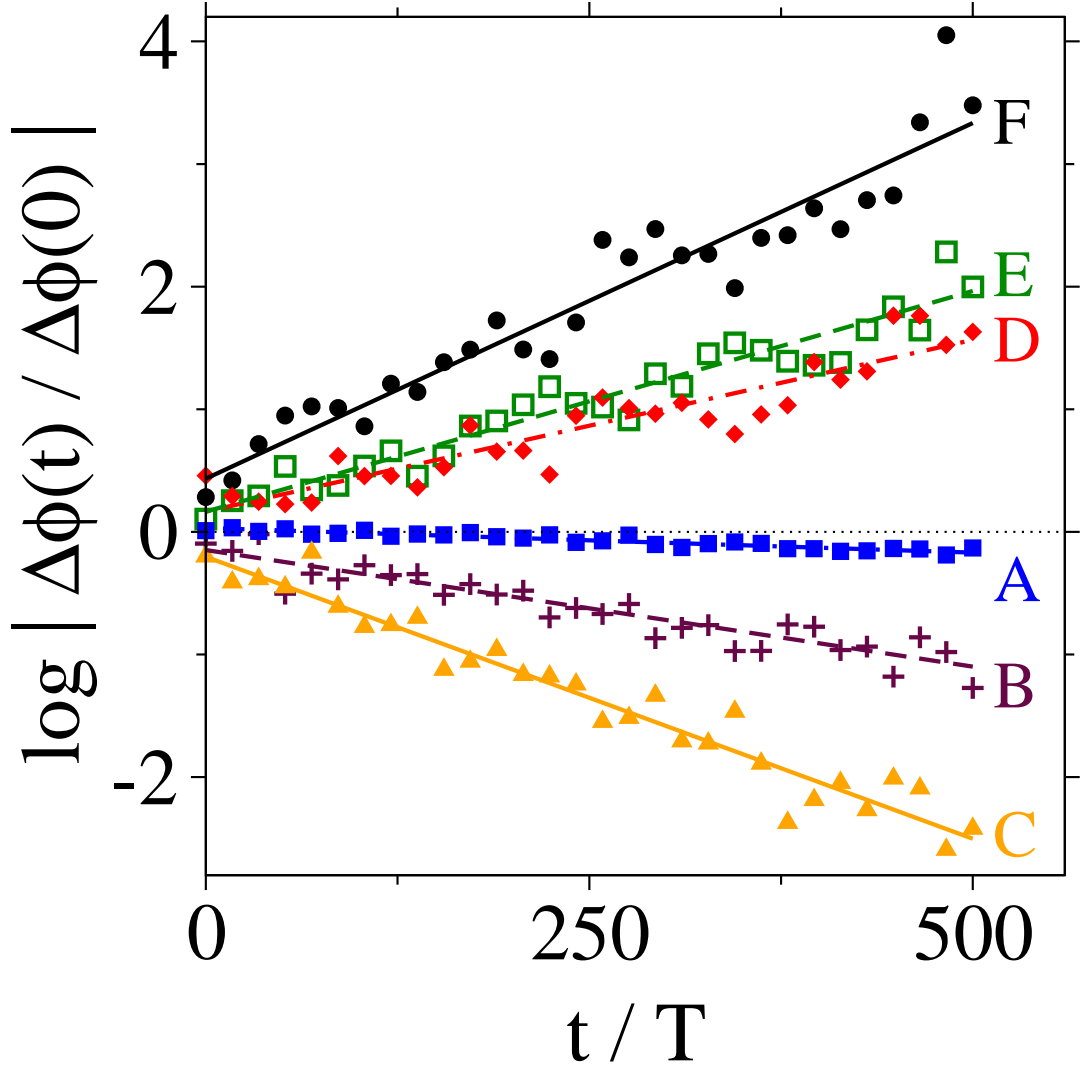}
  \caption{(Color online) 
    Pairwise growth times of perturbations to FHN at $I_0 = 0.34$ for
    several values of the impulse intensity (A: $c=0.03$, B:
    $c=-0.6$, C: $c=0.4$, D: $c=0.2$, E: $c=-0.35$, and F: $c=-0.2$).
    Data was taken for 100-200 trials, each with an ensemble of
    20 oscillators, and with Poisson impulse rate $\lambda = 1/4T$.
    Slope of linear least-square fit gives the Lyapunov exponent.  }
  \label{Fig:FHNMeasuredLyapunov}
\end{figure}

\begin{figure}[htbp]
  \centering
  \includegraphics*[width=0.8\hsize]{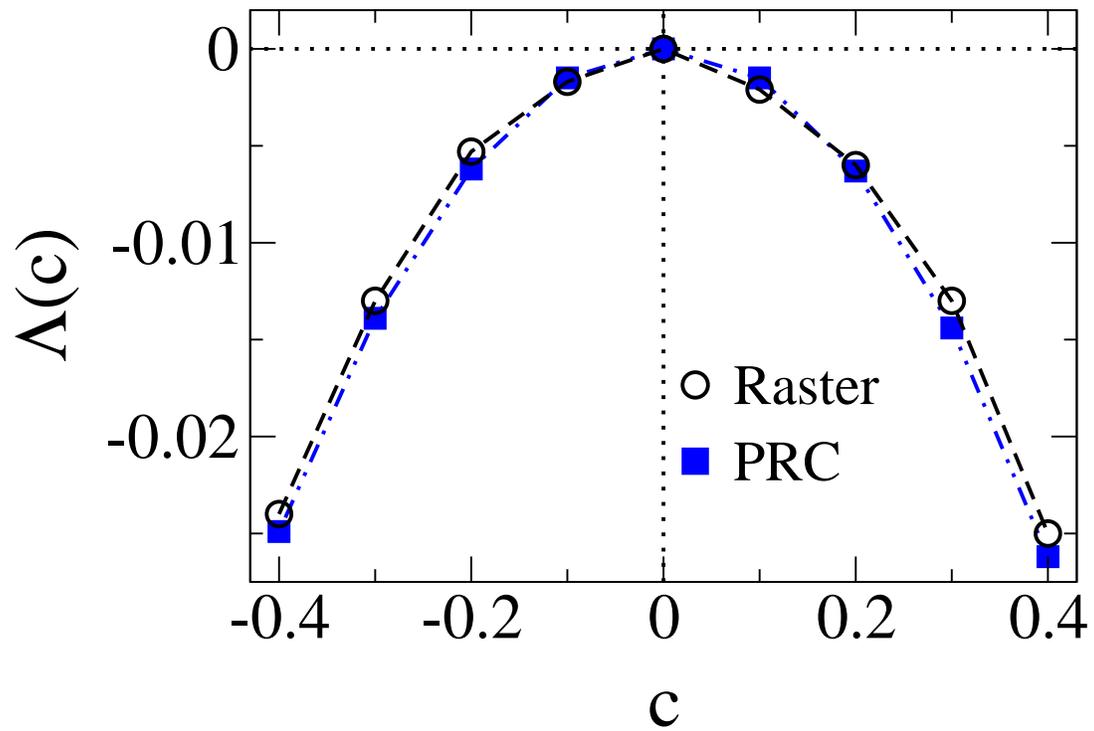}
  \caption{(Color online)  Comparison of the Lyapunov exponents $\Lambda$ between
    the direct measurement from the raster plot and the theoretical
    prediction from the PRC for FHN with parameter $I_0 = 0.8$ driven
    by additive impulses of intensity $c$, and Poisson impulse rate $\lambda = 1/4T$.}
  \label{Fig:SimLyapI0.8}
\end{figure}

\begin{figure}[htbp]
  \centering
  \includegraphics*[width=0.8\hsize]{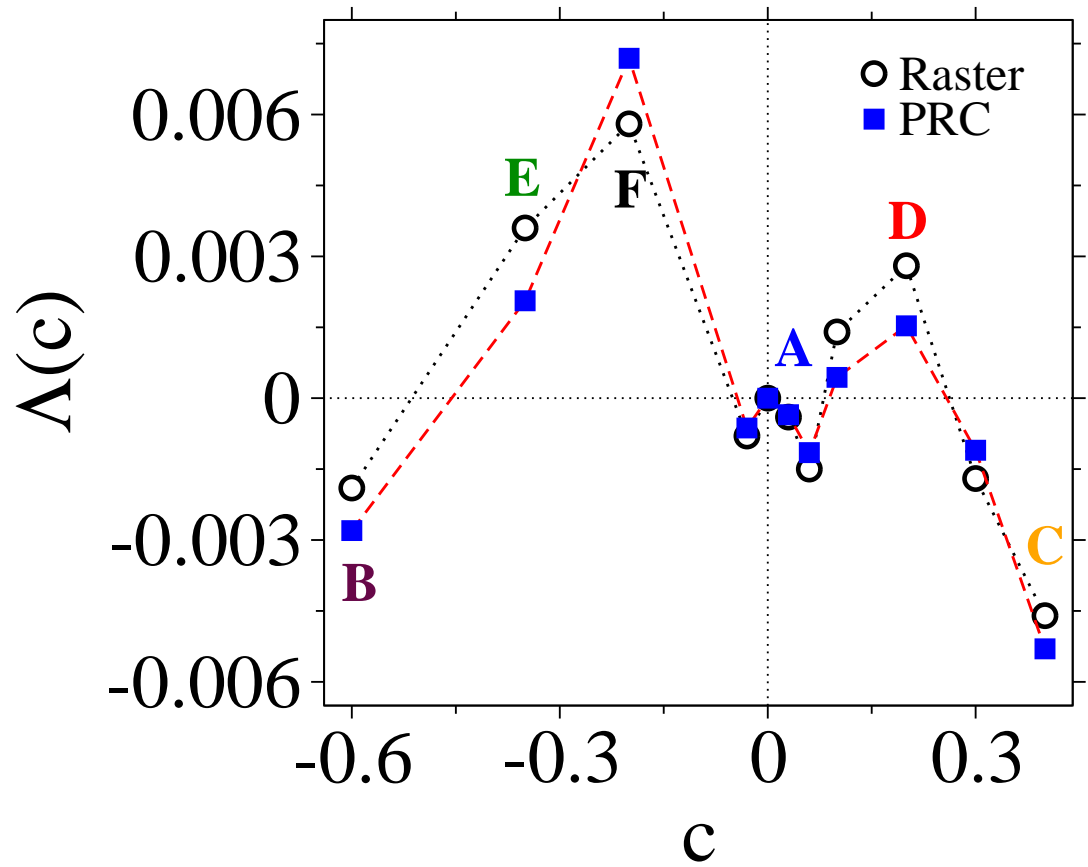}
  \caption{(Color online) Comparison of the Lyapunov exponents $\Lambda$ between
    the direct measurement from the raster plot and the theoretical
    prediction from the PRC for FHN with parameter $I_0 = 0.34$ and
    additive impulses with rate $\lambda = 1/4T$.
    Labels A, B, $\cdots$, $F$ correspond to those in
    Fig.~\ref{Fig:FHNMeasuredLyapunov}.  }
  \label{Fig:SimLyapI0.34}
\end{figure}

To quantitatively test our theory, we measured the Lyapunov exponent
$\Lambda$ in two ways, from numerically obtained PRC $G(\phi, c)$
using Eq.~(\ref{LyapunovFromG}) and by observing the growth rate of
the phase difference of two oscillators with period $T$ using raster
data, Fig.~\ref{Fig:FHNRasterPlots}, for the case of common additive
impulses.
When the time difference $\Delta t$ between the respective $\phi = 0$
events of two oscillators is below a threshold value ($\Delta t < 0.02
T$), we converted the time difference into $\Delta \phi(0) = \Delta t
/ T$, and followed the evolution by measuring subsequent $\Delta
\phi(t)$.
This was done for many oscillator pairs, and for a given $t$, we
calculated the average of $\mbox{log} \left|\Delta \phi(t) / \Delta
  \phi(0)\right|$, as shown in Fig.~\ref{Fig:FHNMeasuredLyapunov} for
$I_0 = 0.34$.  The slope of this line gives $\Lambda$.
Figures~\ref{Fig:SimLyapI0.8} and~\ref{Fig:SimLyapI0.34} show that
Lyapunov exponents measured from the simulation data show good
agreement with those measured from PRCs as shown in
Figs.~\ref{Fig:FHNPRC}(a) and (b), respectively.

\subsection{On-off intermittency and switching}

While the instantaneous growth rate of the difference between two
orbits, $\log | 1 + G'(\phi, \bm{c})|$, may be a smoothly varying
function with respect to phase $\phi$, since the common impulses are
received at random phases, the separation between two oscillators can
be considered a random multiplicative process driven by the
fluctuations in the growth rate around the average Lyapunov
exponent $\Lambda$.
In the absence of any disturbances, complete synchronization
  results if the average effect of a common impulse causes small
    phase perturbations to shrink as indicated by the average Lyapunov
  exponent over the limit cycle, Eq.~(\ref{LyapunovFromG}).
However, if small disturbances exist in the system, fluctuations
  in the instantaneous growth rate can occasionally amplify a
  small deviation in the system, causing intermittent transient
  desynchronization.

\begin{figure}[htbp]
  \centering
  \includegraphics*[width=0.85\hsize]{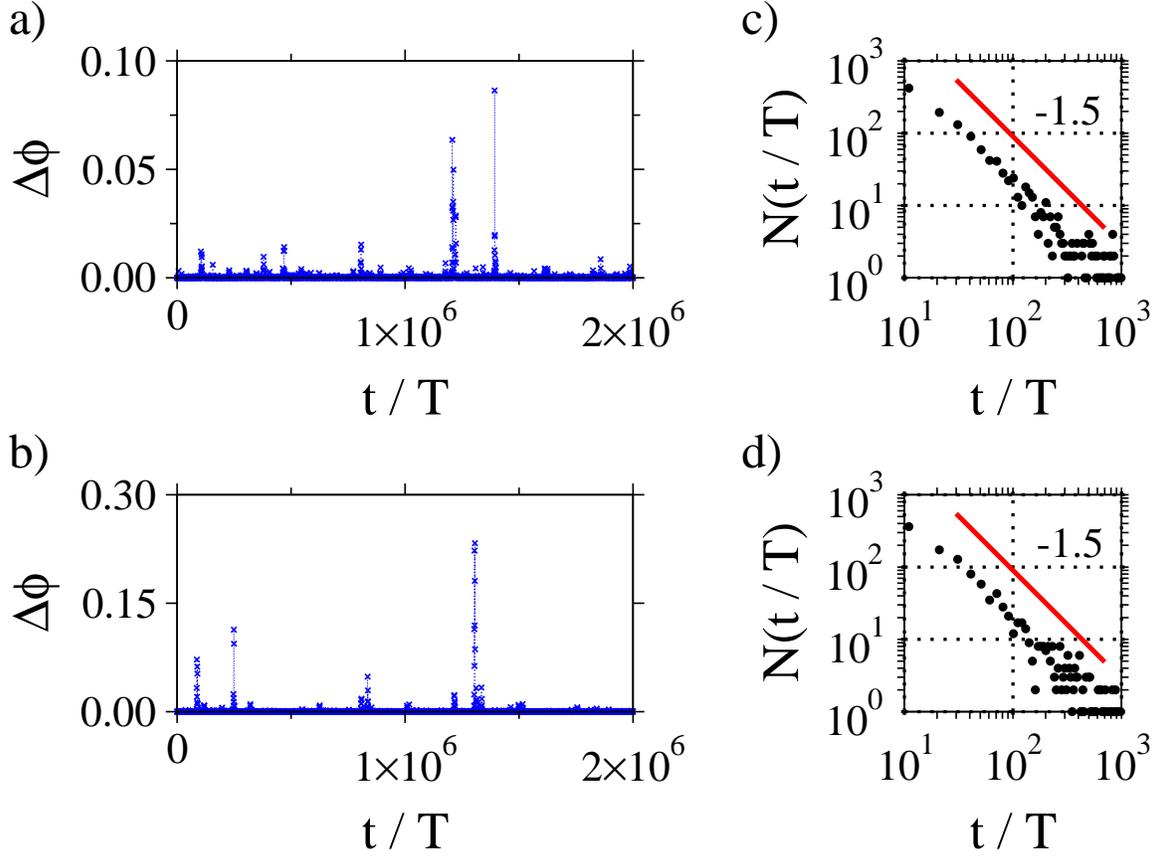}
  \caption{(Color online) On-off intermittency exhibited by 2 oscillators in
    synchronized ($I_0 = 0.8$, additive impulses with $c =
    0.1$), and clustered ($I_0 = 0.875$, linear multiplicative
    impulses with $c = 0.1$) states.
    Independent noise with $D = 9 \times 10^{-9}$, and impulses with $\lambda = 1/4T$ were used.
    Phase difference $\Delta\phi$ between the oscillators from stable
    configuration is small (laminar region) much of the time, but
    large occasional bursts occur.
    (a) long-time evolution of $\Delta\phi(t)$, which shows 
    excursions away from the synchronized state.
    (b) distribution of laminar duration corresponding to (a) (arbitrary normalization).
    (c) long-time evolution of $\Delta\phi(t)$ from the
    $1/2$-out-of-phase clustered state, and
    (d) distribution of laminar duration corresponding to (c) (arbitrary normalization).
    Oscillators are considered to be in the laminar state when $\Delta\phi < 0.0013$
    away from synchronized or clustered states.
    Laminar distributions exhibit power laws with exponent $-1.5$.  At
    this weak independent noise intensity, the phase difference
    between the oscillators takes only either $0$ or $1/2$ depending
    on the initial condition, and switching between the clustered
    state occurs is a very rare event.}
  \label{Fig:Intermittency}
\end{figure}

\begin{figure}[htbp]
  \centering
  \includegraphics*[width=0.85\hsize]{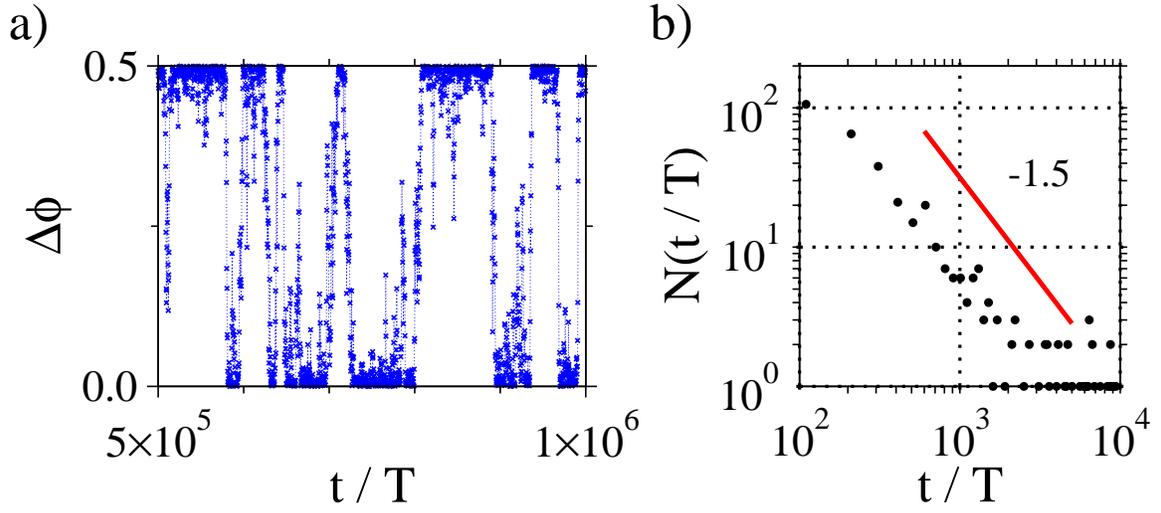}
  \caption{(Color online) Transition between clustered states for 2 oscillators.
    $\Delta\phi$ is the phase difference between the two oscillators.
    Poisson impulses with rate $\lambda = 1/4T$ and $c = 0.1$ was
    used.  A larger independent noise with $D = 3 \times 10^{-4}$ is
    added in order to facilitate the transitions between single and
    2-cluster states.}
  \label{Fig:Switching}
\end{figure}

Figures~\ref{Fig:Intermittency} (a), (c) show an example of the large
excursions away from the synchronized and clustered state
respectively, while Fig.~\ref{Fig:Intermittency}(b), (d) show the
power-law PDF of laminar duration with the well-known exponent,
$-1.5$~\cite{Fujisaka-Yamada,Heagy,Venkaratamani,Pikovsky-R-K1}.
When the system exhibits clustering, the same mechanism leads to
switching for sufficiently strong independent noises, with a
power-law PDF of the life-times of the states as shown in
Fig.~\ref{Fig:Switching}.

This fluctuation is known as modulational or on-off intermittency,
which was first discovered in a pioneering work by Fujisaka and
Yamada~\cite{Fujisaka-Yamada} on a system of coupled chaotic
oscillators, and later clarified to be an ubiquitous feature of many
nonlinear dynamical systems with a symmetry.
The intermittency typically arises in synchronization problems of
dynamical elements, where a dynamical variable acts as a
time-dependent fluctuating driving parameter for a second
variable~\cite{Heagy,Venkaratamani,Pikovsky-R-K1}, for example in the
synchronization of chaotic lasers~\cite{Sauer-Kaiser}, in spin-wave
instabilities~\cite{Roedelsperger}, and in nematic
convection~\cite{John}.
The same mechanism also applies to the present synchronization
  phenomenon induced by common random signals, where the phase
  difference is multiplicatively modulated by rapid random forcing due
  to the common random signals.
  The statistical properties of the separation of trajectories in such
  a situation was previously investigated in detail in
  Ref.~\cite{NoisyPikovsky} for two uncoupled maps receiving a common
  noisy drive, a system very similar to the one currently under
  consideration.

\section{Experiment}

In this section, we present the results of our experiments on an
electrical limit-cycle oscillator.
Synchronization of electrical limit-cycle oscillators induced by
  a continuous random signal has been realized e.g. in
  ~\cite{Yoshida}, but without quantitative comparison with the
  theory.
The deduction of the PRC of noisy neural
  oscillators~\cite{Galan-Ermentrout-Urban} and the use of the PRC in
  predicting oscillation stability, or firing reliability, for cells
  receiving complex stochastic input~\cite{Tateno-Robinson} have also
  been discussed in the neuroscience literature recently, which mainly
  focus on the synchronizing effects of common random signals.
In this section, we experimentally measure the PRC in our
  electric circuit experiments, and quantitatively compare the
  Lyapunov exponent theoretically predicted from the PRC with that
  measured directly from experimental time sequences, in both
  synchronization and desynchronization regimes.

\subsection{Setup}

To observe common impulse-induced synchronization and
desynchronization, we experimented on an electrical limit-cycle
oscillator.
Figure~\ref{Fig:ExptLC} shows the circuit diagram and limit cycle for
our experimental system, a battery-powered LED-flashing oscillator,
where the natural period of the oscillator is $T \approx 0.79$s.
Voltages were measured at two locations in the circuit, $Ch_1$ and
$Ch_2$.  These voltages were taken to be phase space variables with
which we measured the limit cycle.  The limit cycle displayed slight
wandering in the phase space of $V_{ch1} \times V_{ch2}$ as the
natural frequency of the oscillator drifted on the order of $1\%$ over
the course of the experiment, so a new $\phi = 0$ line in phase space
was uniquely chosen every 5 experimental trials after which the limit
cycle was re-calibrated.  This drift in frequency is equivalent to a
small non-identicality of oscillators in an ensemble.  The $\phi = 0$
line was drawn perpendicular to the region of the limit cycle where
the oscillator displayed the fastest dynamics, as such a choice
ensured that the $\phi = 0$ crossing event could be measured with the
least uncertainty.

The impulses were created by computer, which delivered a voltage
signal $V_g$ via the output of a data acquisition card to the
gate of the metal?oxide?semiconductor field-effect transistor (MOSFET) $\mbox{M}_1$ acting as a constant current source, i.e.
a state-independent, additive impulse.
In order to simulate an ensemble of identical oscillators receiving
common impulses, the experiment was repeated many times employing an
identical train of impulses throughout the trials with either random
or identical initial conditions to investigate synchronization or
desynchronization, respectively.

\subsection{Phase response curves}

As in the simulation, we obtained the PRC experimentally by
  applying impulsive stimulus to the circuit.
We varied the PRC characteristics by varying the location of
the circuit to which impulses were applied.
Figure~\ref{Fig:ExptG} shows the experimental PRCs $G_1(\phi,
  V_g)$ and $G_2(\phi, V_g)$ obtained by stimulating $Ch1$ and $Ch2$
  for several impulse intensities.
Qualitatively different PRCs were obtained when we stimulated
  different locations.
In contrast, varying the stimulus intensity to the same location
  resulted in similar but systematically different PRCs.

An experimental measurement has many sources of noise, which seriously
degrades the first derivative calculated from a discretely sampled
PRC, which is necessary in calculating the Lyapunov exponent.
Assuming that the fluctuations seen in the derivative of the PRC are due
to experimental noise and that the underlying response of the
system is smoothly changing with respect to phase, we must infer the
underlying smooth response from our noisy data.

To this end, we smoothed the PRC using a non-Gaussian Kalman
filter~\cite{Kitagawa} based on a Markov state space model with
a transition probability distribution given by the Cauchy
distribution.  The Kalman filter iteratively finds the "true" values,
$ \{\bar{\phi}_1, \bar{\phi}_2, \cdots, \bar{\phi}_N\}$, at each $i$
of the data given the observed data $\{\phi_1, \phi_2, \cdots,
\phi_N\}$ such that conditional probability
\begin{equation}
  p(\{\bar{\phi}_i\} | \{\phi_i\}, \{\mbox{filter parameters}\}),
\end{equation}
is maximized.  The method may be combined with Bayesian methods
such as the Expectation-Maximization algorithm to choose the best
filter parameters.  We did not perform this step, but chose parameters
that preserved the global shape of the PRC while yielding a smooth
first derivative.
This smoothed PRC was then used in Eq.~(\ref{LyapunovFromG}) to
calculate the Lyapunov exponent.

\subsection{Synchronization and desynchronization}

\begin{figure}[htbp]
  \centering
  \includegraphics*[width=0.8\hsize]{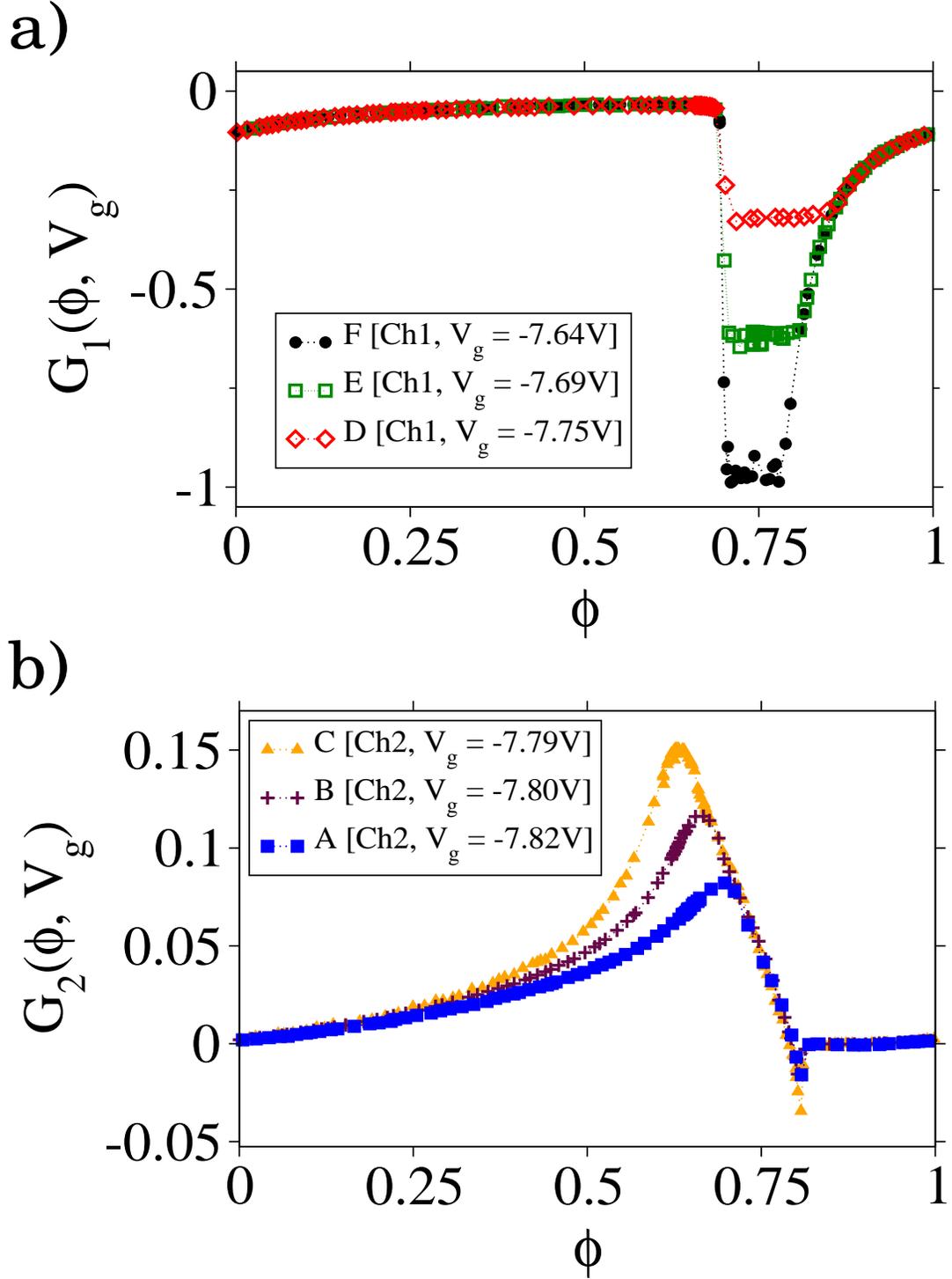}
  \caption{(Color online) PRCs $G_1(\phi, V_g)$ and $G_2(\phi, V_g)$ of electrical
    oscillator obtained by stimulating (a) $Ch_1$, and (b) $Ch_2$,
    which show responses of oscillators that desynchronize and
    synchronize, respectively, upon receiving common impulses.  Each
    curve is labeled with a letter (A, B, C, $\cdots$) that
    corresponds to a location where impulse was applied ($Ch_1$ or
    $Ch_2$), and the MOSFET gate voltage creating the impulse, which
    corresponds to the Poisson mark $c$.}
  \label{Fig:ExptG}
\end{figure}

\begin{figure}[htbp]
  \centering
  \includegraphics*[width=0.8\hsize]{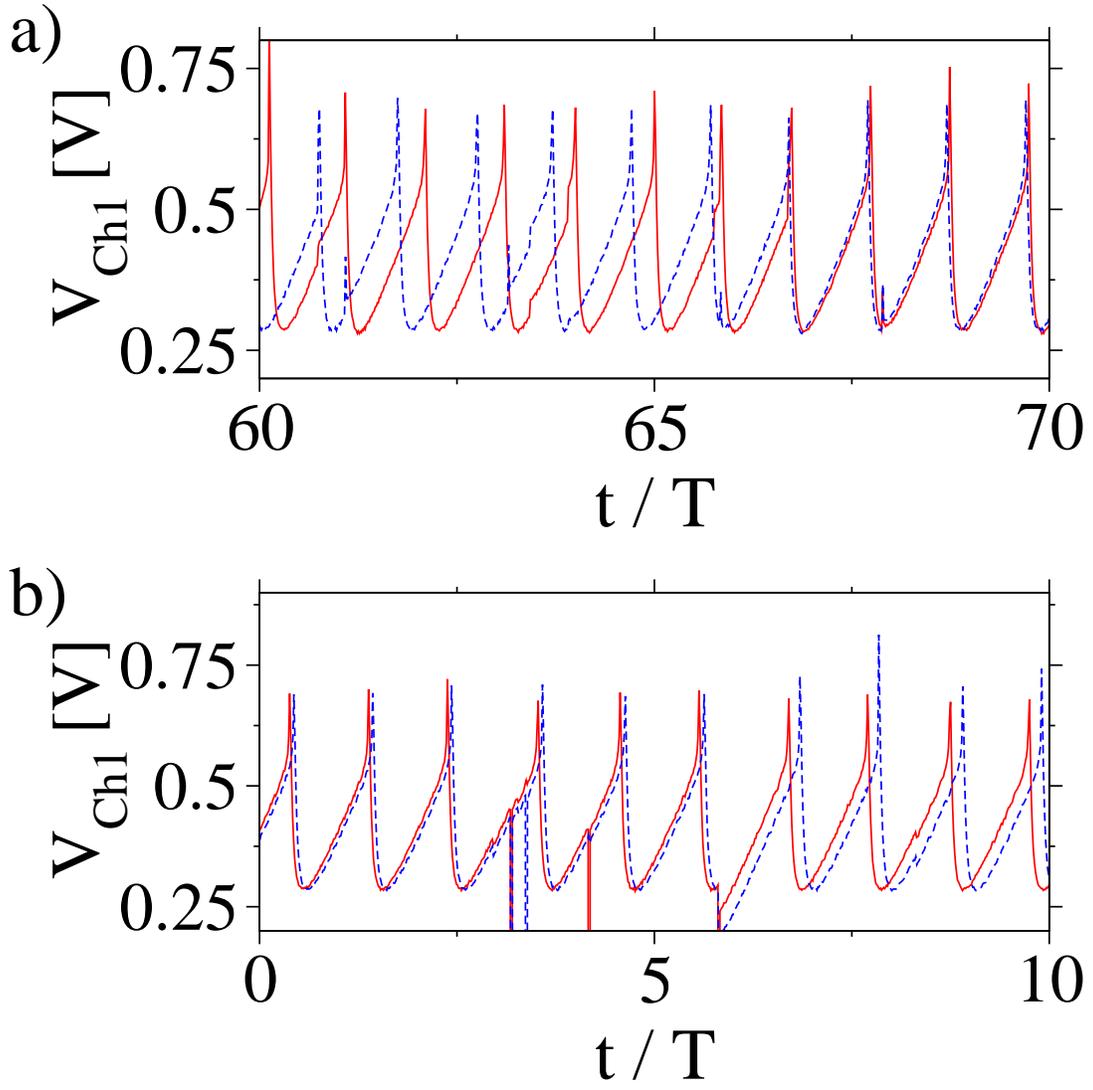}
  \caption{(Color online) Representative waveforms of electrical oscillators
    undergoing common-impulse induced synchronization ($V_g = -7.79$V
    added to $Ch_2$) (a) and desynchronization ($V_g = -7.69$V added
    to $Ch_1$) (b) measured at $Ch_1$.  The Poisson impulse rate is
    $\lambda = 1 / 4T$.  Voltages traces in (b) actually extend below
    $0.2$V, but have been clipped to show detail.}
  \label{Fig:Waveforms}
\end{figure}

\begin{figure}[htbp]
  \centering
  \includegraphics*[width=0.8\hsize]{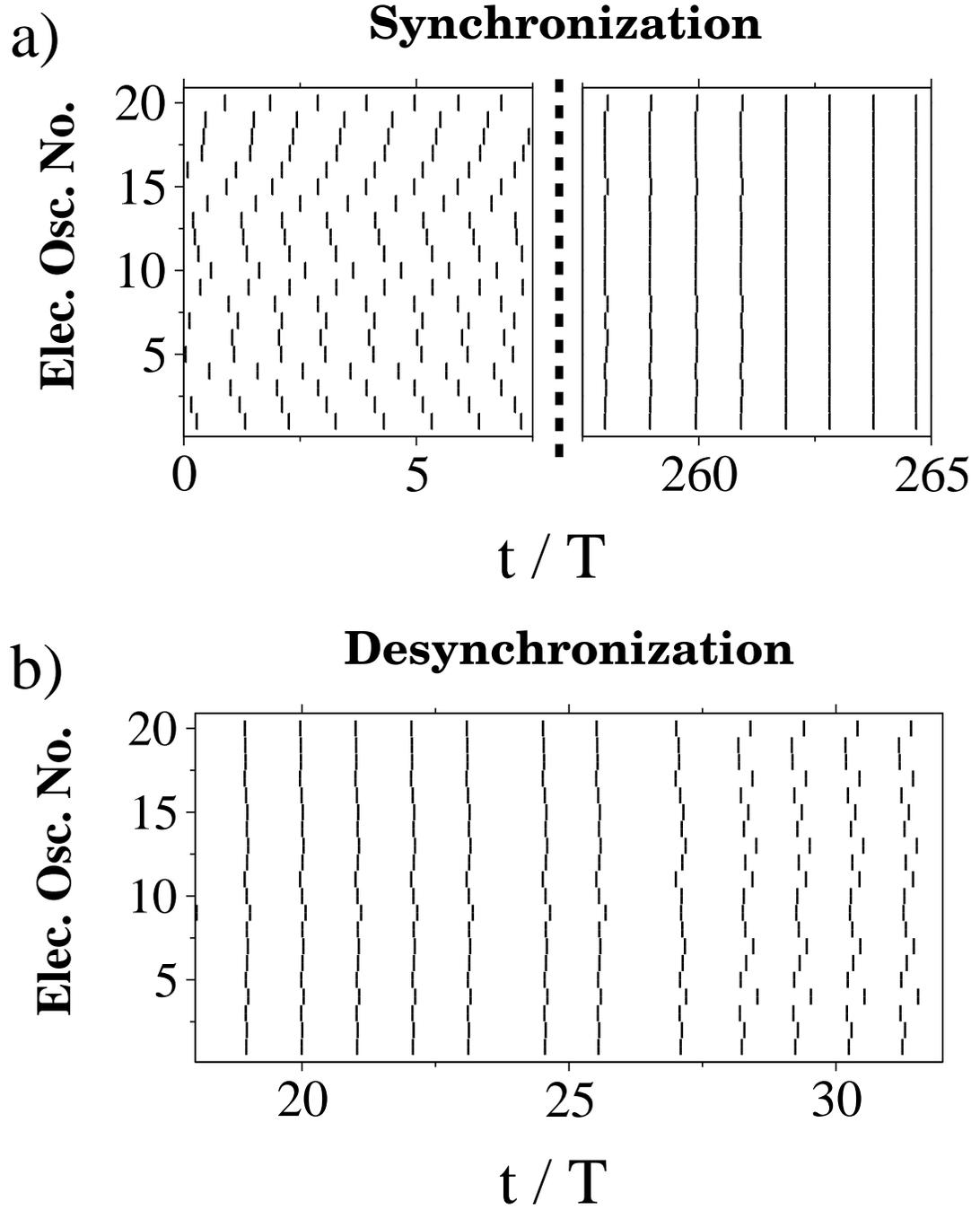}
  \caption{Raster plots of electrical oscillators showing
    synchronization and desynchronization.  The Poisson impulse rate
    is $\lambda = 1 / 4T$.  Each tick indicates the time oscillator
    passed through $\phi = 0$.  (a) Synchronization of electrical
    oscillators (impulse added to $Ch_2$, $V_g = -7.79V$) and (b)
    Desynchronization of electrical oscillators (impulse added to
    $Ch_1$, $V_g = -7.64V$).}
  \label{Fig:ExptRasterPlots}
\end{figure}

\begin{figure}[htbp]
  \centering
  \includegraphics*[width=0.8\hsize]{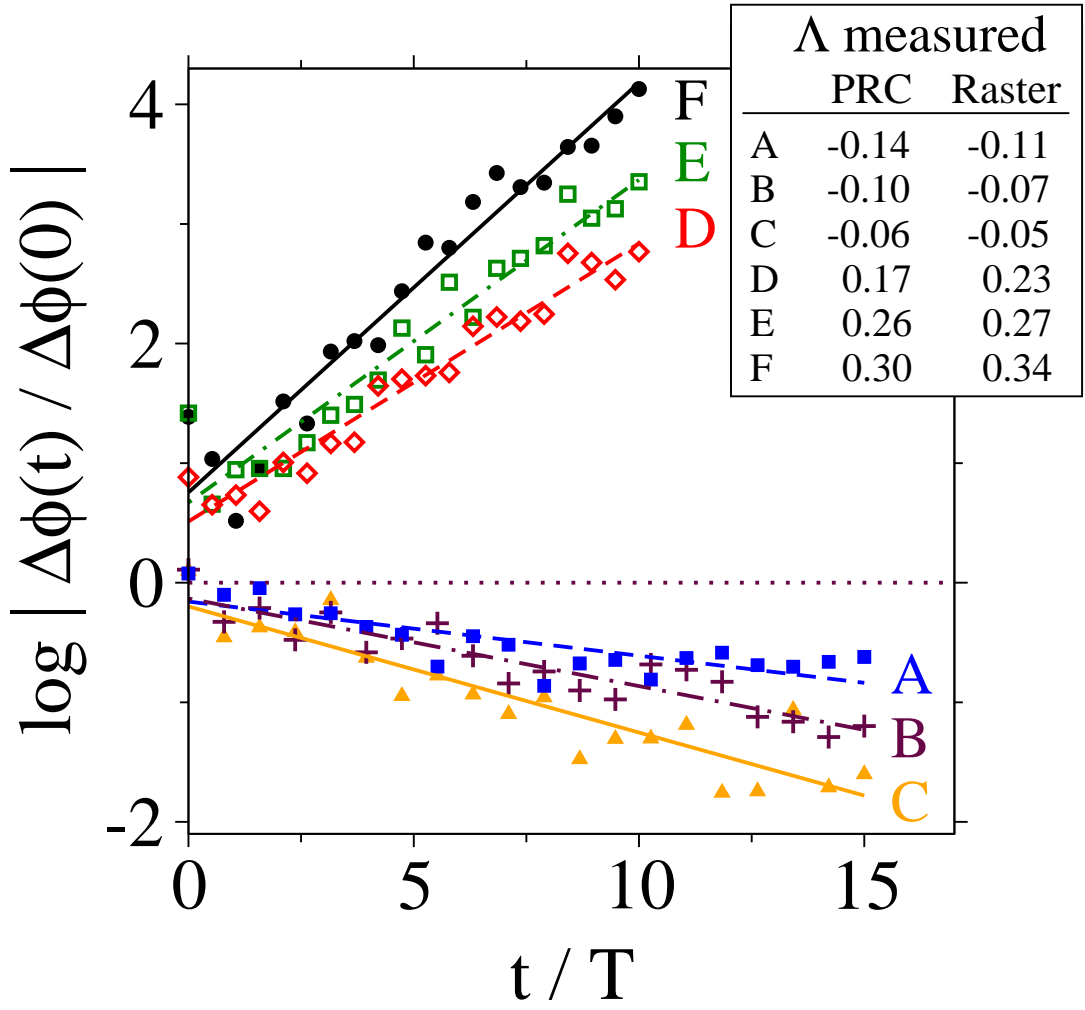}
  \caption{(Color online) 
    Growth times of perturbations to electrical oscillator,
    where (A, B, C, $\cdots$, $F$) correspond to that introduced in Fig.~\ref{Fig:ExptG}.
    Data was taken for 80-150 trials, each with an ensemble of
    20 oscillators, with Poisson impulse rate $\lambda = 1 / 4T$.  
    A comparison of the Lyapunov exponent with theory is shown
      in the table.}  
  \label{Fig:ExpLyap}
\end{figure}

Figure~\ref{Fig:Waveforms} shows the voltage traces at $Ch_1$ for
several different trials showing electrical oscillators synchronizing
and desynchronizing due to common impulses.
The PRC obtained when impulses were applied to $Ch_1$,
Fig.~\ref{Fig:ExptG}(a), had a large jump, and the effect of the
common impulse was predominantly a desynchronizing one.
We were able to detect a synchronization-desynchronization transition
in $Ch_1$ by varying the impulse intensity (data not shown), but
because the synchronizing effects were relatively weak,
we chose instead to investigate the synchronization due to impulses
applied to $Ch_2$, where the PRC is smooth as shown in
  Fig.~\ref{Fig:ExptG}(b) so that we were able to observe
common-impulse induced synchronization clearly, as shown in
  Fig.~\ref{Fig:ExptRasterPlots}(a).
We verified the theoretical predictions by measuring the Lyapunov
exponent using the two methods outlined in the previous
  section. The results of the two methods of measurement of the
Lyapunov exponent are summarized in the caption of
Fig.~\ref{Fig:ExpLyap}.
Considering the frequency drift among trials, the agreement between
the results of the two methods seems reasonable.
\footnote{The average log ratio near $t = 0$ for the
  desynchronization times do not go to $0$ as expected.  This is due
  to the slight ($\approx 1\%$) drifting in the frequency exhibited by
  the oscillators throughout the experiment.  This non-ideality affects
  the desynchronization and synchronization times differently due to
  the fact that the mismatch causes accelerated
  synchronization and desynchronization (canceling out on average)
  when measuring synchronization times, while it only causes premature
  desynchronization when measuring desynchronization. }.

Note that our electrical limit-cycle oscillator is not near the
  bifurcation point and the PRC is not jagged in contrast to the FHN
  oscillator, even though the oscillators are desynchronized when
  impulses are applied.  Therefore, the desynchronization effect
  is rather gradual, as shown in the raster plot,
  Fig.~\ref{Fig:ExptRasterPlots}(b), which is qualitatively similar to
  the situation with strong impulses on the Stuart-Landau oscillators
  that we discuss in Appendix E.

\section{Summary}

Through theoretical analysis, numerical simulations, and circuit
experiments, we have demonstrated that phase synchronization,
desynchronization, and clustering can be realized in a system of
identical, uncoupled oscillators acted on by common Poisson impulses
by changing the parameters of each individual oscillator, and also by
varying the strength of the common impulse.
We have clarified that once the phase response curves of the
oscillator is known, such phase coherence phenomena can be
quantitatively understood in a unified way.
The desynchronization is an effect of the finite size of impulses that
we use, in sharp contrast to the exclusively synchronizing effects
that infinitesimal perturbations have.

It is quite remarkable that the synchronization and desynchronization
of an ensemble of oscillators may be controlled in this way through a
random signal.  Depending on the oscillator characteristics, we could
alter the coherence properties of the ensemble simply by changing the
intensity of external random impulses.  This approach would have
potential applications in which it is desirable to change the global
coherence properties of a network of oscillators.  Applying common
impulse of suitable intensity, it would be possible to effectively
"switch off" or "switch on" the synchronization without changing the
oscillator characteristics or modifying the coupling strength.

In this paper, we characterized the local stability of the
synchronized states by a Lyapunov analysis, but not the global
stability.  In Ref.~\cite{Nakao-Arai-Kawamura}, we have presented a
global stability analysis of the system for common Gaussian-white
driving using an averaging technique for nonlinear oscillators.
Similar formulation is also possible for the present common impulsive
driving.
The clustering phenomenon was realized only numerically for the FHN
system, because our present experimental circuit does not appear to
have a symmetric limit cycle or PRC compatible with clustering.
Other limit-cycle electric oscillators with nearly symmetric limit
cycles do exist, and investigation of clustering with such systems is
now under progress.
We expect that on-off intermittency and switching can also be observed
in such electric oscillators.
Detailed analysis on these topics will be reported in the future.

\section{Acknowledgments}

The authors wish to thank Y. Kuramoto for helpful discussions,
especially suggesting us to analyze the impulse-driven case, and H.
Fujisaka for all his insightful comments and advice.
We also thank K. Aihara, A. Uchida, K.  Yoshimura, H. Suetani, T. Shimokawa, J.
Teramae, D.  Tanaka, T.  Kobayashi, Y. Kawamura and Y. Tsubo for various useful
comments and information.
This work is supported partially by the Grant-in-Aid for the 21st
Century COE "Center for Diversity and Universality in Physics", and
partially by the Grant-in-Aid for Young Scientists (B), 19760253,
2007, from the Ministry of Education, Culture, Sports, Science, and
Technology of Japan.


\appendix

\section{Poisson-driven Markov process (Jump process)}

In this paper, we model the process of an oscillator receiving
impulses as a Poisson-driven Markov process, or {\em jump
  process}~\cite{Snyder,Hanson}.
A general stochastic process $\bm{X}(t)$ driven by Poisson random
impulses,
\begin{equation}
  \dot{\bm{X}}(t) = \bm{F}(\bm{X})
  + \sum_{n=1}^{N(t)} \bm{G}(\bm{X}, \bm{c}_n) h(t - t_n)
\end{equation}
 with point-like impulses $h(t)$ should be interpreted in the Ito
  picture as
\begin{equation}
  \bm{X}(t)
  =
  \bm{X}(0) + \int_0^t \bm{F}(\bm{X}(s)) ds
  +
  \sum_{n=1}^{N(t)} \bm{G}(\bm{X}(t_n - 0), \bm{c}_n),
\end{equation}
which is described as an integral equation of the form
\begin{equation}
  \bm{X}(t)
  =
  \bm{X}(0) + \int_0^t \bm{F}(\bm{X}(s)) ds
  +
  \int_0^t \int_{\bm{c}} \bm{G}(\bm{X}(s), \bm{c}) M(dt, d\bm{c}),
\end{equation}
or as a jump stochastic differential equation (SDE),
\begin{equation}
  d\bm{X}(t) = \bm{F}(\bm{X}) dt + \int_{\bm{c}} \bm{G}(\bm{X}, \bm{c}) M(dt, d\bm{c}).
\end{equation}
Here, $M(dt, d\bm{c})$ represents a Poisson random
measure~\cite{Snyder,Hanson}, which gives the number of incident
points during $[t, t+dt]$ having mark (jump magnitude) in
$[\bm{c}, \bm{c}+d\bm{c}]$, and $N(t) = \int_0^t \int_{\bm{c}} M(dt,
d{\bm{c}})$ is the number of incident points during $[0, t]$.
As usual in Ito-type stochastic differential equations,
  functions of ${\bm X}(t)$ and the Poisson random measure $M(dt,
  d\bm{c})$ at the same instant of time are independent.
The expectation of $M(dt, d\bm{c})$ is
\begin{equation}
  \langle M(dt, d\bm{c}) \rangle = \lambda dt p(\bm{c}) d\bm{c},
\end{equation}
where $\lambda$ is the rate of the Poisson process and the marks are
distributed as $p(\bm{c})$.
Similarly, the covariance of $M(dt_1, d\bm{c}_1)$ and $M(dt_2,
d\bm{c}_2)$ is~\cite{Snyder,Hanson}
\begin{eqnarray}
  \mbox{Cov}[M(dt_1, d\bm{c}_1), M(dt_2, d\bm{c}_2)] 
  =
  \lambda \delta(t_2 - t_1) dt_1 dt_2 p(\bm{c}_1) \delta(\bm{c}_1 - \bm{c}_2) d\bm{c}_1 d\bm{c}_2.
\end{eqnarray}
As in the case of Wiener-driven Ito stochastic differential
  equations, we need to use a special rule in calculating the
  differential of a transformed process.  A transformed process $V(t)
  = V(\bm{X}(t))$ obeys a jump SDE of the form
\begin{equation}
  dV(t) = \left[ \mbox{grad}_{\bm{X}} V(\bm{X}) \cdot \bm{F}(\bm{X}) \right] dt +
  \int_{\bm{c}} [V(\bm{X} + \bm{G}(\bm{X}, \bm{c})) - V(\bm{X})] M(dt, d\bm{c}),
\end{equation}
which is a stochastic chain rule for the jump process.
The chain
rule changes an additive process into a multiplicative process, just as in the case of white-Gaussian driven Markov processes.

\section{Single oscillator phase distribution}

\begin{figure}[htbp]
  \centering
  \includegraphics*[width=0.8\hsize]{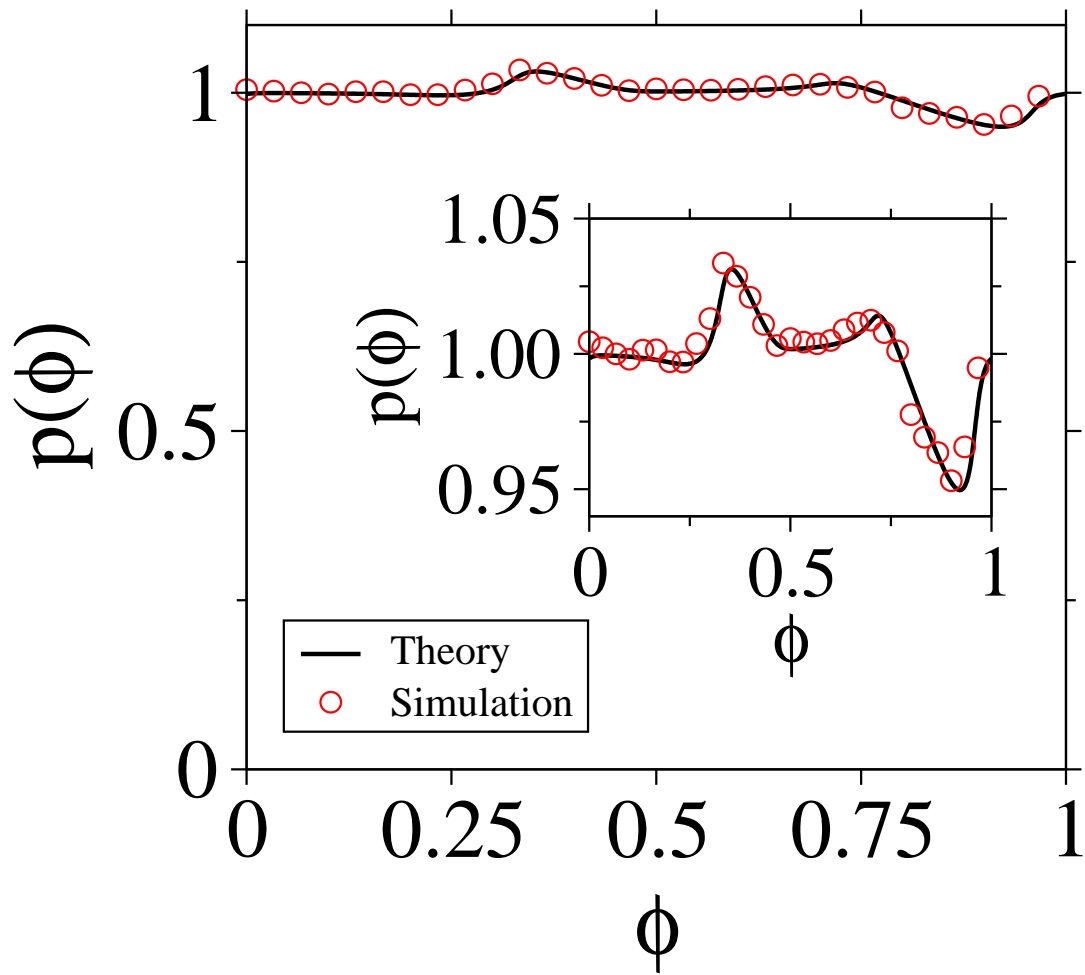}
  \caption{(Color online) The stationary phase distribution of a FitzHugh-Nagumo
    oscillator receiving random Poisson impulses, $I_0 = 0.8$, $c =
    0.5$ and $\lambda = 1 / 4T$, calculated using a
    perturbation expansion of the forward Kolmogorov equation and by
    direct numerical simulation.}
  \label{Fig:1OscPhaseDist}
\end{figure}

In the calculation of the Lyapunov exponent, we simplified the
calculation by assuming that a single oscillator receiving random
Poisson impulses is evenly distributed in phase when averaged over a
long period of time.  This is strictly not the case, as can be seen
from the map $\phi(\theta) = \theta + G(\theta, {\bm{c}})$.
It is obvious that certain values of $\phi(\theta)$ are arrived at
more often than other values, because $G(\theta, \bm{c})$ is
  generally a nonlinear function of $\theta$.  For
  completeness, we calculate the deviation from a flat phase PDF by
the forward Kolmogorov equation~\cite{Hanson}.
To this end, we define $\eta(\phi, {\bm{c}}) = G(\theta, {\bm{c}})$,
which is the jump as a function of the destination coordinate $\phi$.
Then the probability density $p(\phi, t)$ of $\phi$ obeys
\begin{eqnarray}
  \pderivop{t}p(\phi, t) = 
  -\pderivop{\phi}[\omega p(\phi, t)] - \lambda p(\phi, t) 
  + \lambda \int_{\bm{c}} p({\bm{c}}) p(\phi - \eta(\phi, {\bm{c}}))
  \left| 1 - \pderivop{\phi} \eta(\phi, {\bm{c}}) \right| d{\bm{c}}.
\end{eqnarray}
We find the stationary PDF by setting $\partial p(\phi, t) / \partial
t= 0$.
It is clear that the combination $\epsilon = \lambda / \omega$
  determines the effect of the impulses, which is a small parameter
from our initial assumption (the rate of the Poisson process
  $\lambda$ is small).
We thus expand $p(\phi)$ in powers of $\epsilon$ away from the
stationary PDF as $p(\phi) = 1 + \epsilon p_1(\phi) + \cdots$.
Substituting this into the forward Kolmogorov equation, the first
non-vanishing terms are of order $O(\epsilon^1)$:
\begin{equation}
  0 = \pderivop{\phi} p_1(\phi) + 1 - 
  \int_{\bm{c}} p({\bm{c}}) \left| 1 - \pderivop{\phi} \eta(\phi, {\bm{c}}) \right| d{\bm{c}}.
\end{equation}
If the PRC does not change too rapidly, $-1 < {\partial \eta(\phi, {\bm{c}})}
/ {\partial \phi} < 1$, and the absolute value of the integrand may be
removed, yielding
\begin{equation}
  \pderivop{\phi} p_1(\phi) = 
  -\int_{\bm{c}} p({\bm{c}}) \pderivop{\phi} \eta(\phi, {\bm{c}}) d{\bm{c}}.
\end{equation}
Then
\begin{equation}
  p_1(\phi) = \mbox{C} -\int_{\bm{c}} p({\bm{c}}) \eta(\phi, {\bm{c}}) d{\bm{c}}.
\end{equation}
Since the integral of $p(\phi)$ over one period of the $O(\epsilon^0)$
term is $1$ due to normalization, higher order terms
$p_1(\phi), p_2(\phi), \cdots$ must vanish upon integration over a
full period.  Specifically, the $\int_0^1 p_1(\phi) d\phi = 0$
condition yields
\begin{equation}
  C = \int_{\bm{c}} p({\bm{c}}) \overline{\eta}({\bm{c}}) d\bm{c}
\end{equation}
where
\begin{equation}
  \overline{\eta}({\bm{c}}) = \int_{0}^{1} \eta(\phi, {\bm{c}}) d\phi.
\end{equation}
Therefore, we have for the first order approximation to the stationary
phase PDF,
\begin{equation}
  p(\phi) = 1 + \epsilon \int_{\bm{c}}
  \left[ \overline{\eta}({\bm{c}}) - \eta(\phi, {\bm{c}}) \right] p({\bm{c}}) d{\bm{c}} + O(\epsilon^2).
\end{equation}
As we see from Fig.~\ref{Fig:1OscPhaseDist}, $p(\phi)$ is close to 
constant as long as the parameter $\epsilon$ is small.
In Ref.~\cite{Nakao-Arai}, we argued that for Poisson impulses, the
lowest order correction to the uniform phase density does not
contribute to the Lyapunov exponent based on a perturbation expansion
of a Frobenius-Perron-type equation.  Here we only point out that the
correction to the uniform density is of $O(\lambda / \omega)$, which
is small if the Poisson rate $\lambda$ is sufficiently smaller than
the oscillator natural frequency $\omega$.

\section{Geometric Interpretation of the Synchronizing Mechanism}

Weak impulses always synchronize uncoupled oscillators.  Here we show
through a simple Floquet analysis that this synchronization is a
consequence of the stability of the limit cycle against weak
perturbations.
The generic oscillator under consideration in this article is
described by an ordinary differential equation of the form
\begin{equation}\label{GenericODE}
  \dot{\bm{X}}(t) = \bm{F}(\bm{X}),
\end{equation}
with a stable limit-cycle solution $\bm{X}_0(t)$ with period $T$.
Linearizing Eq.~(\ref{GenericODE}) with respect to a small
perturbation $\bm{u}(t)$ from the limit cycle, we obtain
\begin{equation}
  \dot{\bm{u}}(t) = \bm{D F}(\bm{X})|_{\bm{X} = \bm{X}_0(t)} \bm{u}
\end{equation}
where $\bm{D F}(\bm{X})|_{\bm{X} = \bm{X}_0(t)}$ is a periodic $M
\times M$ Jacobian matrix.  Ordinary differential equations of this
form with periodic coefficients have solutions of the form $\bm{u}(t)
= \bm{Q}(t) e^{\bm{R}t} \bm{u}(0)$ where $\bm{R}$ is a constant $M
\times M$ matrix, and $\bm{Q}(t) = \bm{Q}(t + T)$ is a $T$-periodic $M
\times M$ matrix.  Since $\bm{Q}$ is periodic, we have $\bm{u}(T) =
e^{\bm{R}T}\bm{u}(0)$.  There is a corresponding value of the constant
matrix $\bm{R}$ for each point on the limit cycle.
The eigenvalues and eigenvectors of $e^{\bm{R}T}$, $\{\lambda_i\}$ and
$\{\bm{e}_i\}$ respectively, have the property that $\bm{e}_1$ is in
the direction along the limit cycle, $\lambda_1 = 1$ and $|\lambda_i|
< 1$ for $i \in \{2, \cdots, M\}$.  The eigenvectors are known as the
Floquet eigenvectors, and each point on the limit cycle has its own
set of Floquet eigenvalues and eigenvectors.

\begin{figure}[htbp]
  \centering
  \includegraphics*[width=0.8\hsize]{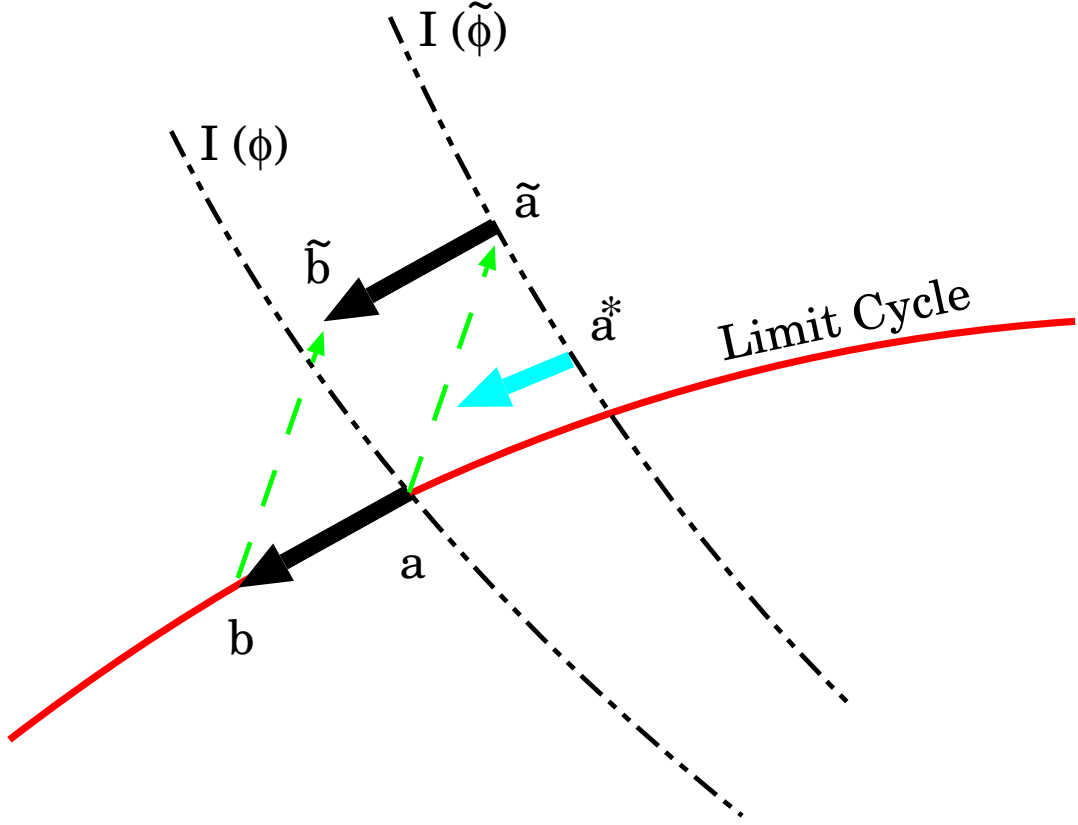}
  \caption{(Color online) Schematic of evolution two nearby orbits.  $a =
    \bm{X}_0(\phi)$ and $b = \bm{X}_0(\phi) + \bm{z}(0)$ represent the
    spatial points at which 2 orbits receive a common additive
    impulse $\bm{c}$ on the limit cycle.  The oscillators jump to
    $\tilde{a} = \bm{X}_0(\phi) + \bm{c}$ and $\tilde{b} =
    \bm{X}_0(\phi) + \bm{z}(0) + \bm{c}$.  After the oscillator
    completes one period of unperturbed motion, through analysis of
    the Floquet eigenvectors and eigenvalues, we see that the
    difference vector has shrunk, i.e.  $\left|\bm{z}(T)\right| <
    \left|\bm{z}(0)\right|$.}
  \label{Fig:Floquet}
\end{figure}

Figure~\ref{Fig:Floquet} shows two infinitesimally separated orbits, 1
and 2, on the limit cycle at $a$, $\bm{X}_0(\phi)$, and
$b$, $\bm{X}_0(\phi) + \bm{z}(0)$, with a phase $\phi$ isochron
passing through $a$.
Now let the two oscillators receive a common additive impulse $\bm{c}$
at $t = 0$.  The two oscillators jump discontinuously to $\tilde{a}$
at $\bm{X}_0(\phi) + \bm{c}$ and $\tilde{b}$ at $\bm{X}_0(\phi) +
\bm{z}(0) + \bm{c}$, with a phase $\tilde{\phi}$ isochron passes
through $\tilde{a}$.
The set of Floquet eigenvalues and eigenvectors at $\phi$ and
$\tilde{\phi}$ are $(\{\lambda_{i}\}, \{\bm{e}_{i}\})$ and
$(\{\tilde{\lambda}_{i}\}, \{\tilde{\bm{e}}_i\})$,
respectively.
Since the impulse is additive, $\stackrel{\rightarrow}{ab} =
\stackrel{\rightarrow}{\tilde{a}\tilde{b}} = \bm{z}(0)$.
Expanding the difference vector $\bm{z}(0)$ by the Floquet vectors, we
obtain $\bm{z}(0) = a_1 \bm{e}_1 = \sum_i \tilde{a}_{i}
\tilde{\bm{e}}_{i}$.  It is then obvious that $|a_1| > |\tilde{a}_1|$.

If the oscillator continues unperturbed for one period, the oscillator
that received the impulse at $a$ at $t = 0$ will now be at $a^*$, and
$\bm{z}(T) \cong e^{\bm{R}T} \bm{z}(0) = \sum_i \tilde{\lambda}_i
\tilde{a}_{i} \tilde{\bm{e}}_{i}$.  Since $\left|\tilde{\lambda}_i\right| < 1$ for $i \in \{2, \cdots,
M\}$ from the stability of limit cycles, all components of
$\bm{z}(T)$ with $i \geq 2$ will shrink, leaving only the first
component along the limit cycle.
We thus see that $|\bm{z}(T)| < |\bm{z}(0)|$ by virtue of
$|\tilde{a}_1| < |a_1|$, namely that the application of a common impulse
always shrinks the small separation between two orbits.

\section{Diffusion limit}

We here derive the diffusion (Gaussian-white) limit of the
  impulse-driven oscillators for weak and frequent impulses.
In taking the diffusion limit, we see that common impulse always
results in the synchronization of oscillators since the Lyapunov
exponent is bounded above by 0, and we see that desynchronization can
only be understood by analyzing finite-magnitude perturbations to the
limit cycle orbit.

\subsection{Diffusion limit}

We have until now considered finite Poisson impulses whose
inter-impulse times are much longer than the natural period of the
oscillators.
 The condition for the phase reduction is also satisfied by
  setting the combination of impulse intensity and the Poisson rate
  appropriately small, so that we may also consider a situation where
  the effect of Poisson impulses become infinitesimal but with
  inter-impulse times that are much faster than the natural time scales
  of the oscillators.
Specifically, we consider the limit $\lambda \to \infty$ and the
  effect of impulses $\sigma_k \to 0$ such that $\lambda \langle
  \sigma_k \sigma_l \rangle_{\bm{c}}$ is kept constant and higher
  order terms like $\lambda \langle\sigma_k \sigma_l \sigma_m\rangle$ vanish
  $(k, l, m = 1, \cdots, M)$.
To take this limit, it is necessary that the net effect of the
  impulses vanishes, namely,
\begin{eqnarray}
  \langle \sigma_k(\phi, \bm{c}) \rangle_{\bm{c}}
  = \int_{\bm{c}} p(\bm{c}) \sigma_k(\bm{X}_0(\phi), \bm{c}) d\bm{c} = 0,
  \;\;\;\;
  (k=1, \cdots, M),
\end{eqnarray}
where we introduced the notation $\langle A(\phi, \bm{c})
\rangle_{\bm{c}} = \int_{\bm{c}} A(\phi, \bm{c}) p(\bm{c})
d\bm{c}$ with fixed $\phi$.
Under these conditions, we can consider the diffusion limit of a
stochastic jump process.

The Kramers-Moyal expansion of the Chapman-Kolmogorov equation
is~\cite{Risken}
\begin{equation}
  \pderiv{p(\phi, t)}{t} = \sum_{n=1}^{\infty} \frac{1}{n!}
  \left( - \frac{\partial}{\partial \phi} \right)^{n}
  \left[ K^{(n)}(\phi) p(\phi, t) \right],
\end{equation}
where the Kramers-Moyal coefficient $K^{(n)}$ is given by
\begin{equation}
  K^{(n)}(\phi) = \lim_{\Delta t \to 0} 
  \frac{\langle \Delta \phi^n \rangle}{\Delta t}.
\end{equation}
Here, $\Delta \phi$ is the jump within duration $\Delta t$ of
  the stochastic process starting from $\phi$, and the conditional
  average is taken over possible realizations of the stochastic
  process starting from $\phi$.
If the coefficients higher than the second order vanish, we
  obtain a Fokker-Planck equation
\begin{equation}\label{DiffLimitFPE}
  \pderiv{p(\phi, t)}{t} = -\pderivop{\phi} \left[ v(\phi) p(\phi, t) \right] + 
  \frac{1}{2} \pnderivop{\phi}{2} \left[ D(\phi)p(\phi, t) \right]
\end{equation}
for the Wiener-driven Markov process, whose drift $v(\phi)$ and
  diffusion coefficient $D(\phi)$ are given by
\begin{equation}
  v(\phi) = K^{(1)}(\phi), \;\;\;\;\; D(\phi) = K^{(2)}(\phi).
\end{equation}
We now find $\langle \Delta \phi \rangle$ and $\langle \Delta
\phi^2 \rangle$ for a jump process, where the expectation is
to be taken with fixed $\phi$.
Starting with Eq.~(\ref{PhaseSIDE}),
\begin{eqnarray}
  \langle \Delta\phi\rangle
  &=&
  \omega \Delta t
  +
  \int_{0}^{\Delta t} \int_{\bm{c}} G(\phi, \bm{c})
  \langle M(dt, d\bm{c}) \rangle
  +
  O(\Delta t^2) \cr \cr
  &=&
  \omega \Delta t + \lambda \Delta t \langle G(\phi, \bm{c}) \rangle_{\bm{c}}
  +
  O(\Delta t^2).
\end{eqnarray}
Similarly,
\begin{eqnarray}
  \langle \Delta\phi^2 \rangle
  &=&
  \langle (\Delta\phi - \langle \Delta\phi \rangle)^2 \rangle + O(\Delta t^2) \cr \cr
  &=&
  \int_0^{\Delta t} \int_0^{\Delta t} \int_{\bm{c}} \int_{\bm{c}'} G(\phi, \bm{c}) G(\phi', \bm{c}') 
  \; \mbox{Cov} \left[ M(dt, \bm{c}), M(dt', \bm{c}') \right] + O(\Delta t^2) \cr \cr
  &=&
  \lambda \Delta t \langle G(\phi, \bm{c})^2  \rangle_{\bm{c}} + O(\Delta t^2).
\end{eqnarray}
It can be checked that $\langle \Delta \phi^3 \rangle$
  and higher-order moments vanish by taking the diffusion limit, so
  that they can be dropped.
The Kramers-Moyal coefficients are given by
\begin{eqnarray}
  K^{(1)}(\phi) = \omega + \langle G(\phi, {\bm c}) \rangle_{\bm{c}}, \;\;\;\;\;
  K^{(2)}(\phi) = \langle G(\phi, {\bm c})^2 \rangle_{\bm{c}}, \;\;\;\;\;
  K^{(n)}(\phi) = 0 \;\; (n \geq 3).
\end{eqnarray}
 Hereafter, for simplicity, we may not explicitly indicate the
  dependence of the function $\bm{g}(\bm{X}_0(\phi), \bm{c})$,
  $\bm{\sigma}(\bm{X}_0(\phi))$, or $\bm{Z}(\phi)$ on $\phi$ and
  $\bm{c}$.
To evaluate the $v(\phi)$ and $D(\phi)$, we must evaluate $\langle
G(\phi, \bm{c}) \rangle_{\bm{c}}$ and $\langle G(\phi, \bm{c})^2
\rangle_{\bm{c}}$.
Since we assume the effect of impulses $\bm{g}$ to be small, we
first rewrite $G(\phi, \bm{c})$ by Taylor expanding,
\begin{eqnarray}
  \label{TaylorG}
  G(\phi, \bm{c})
  & = &
  \phi\bigl(\bm{X}_{0}(\phi) + \bm{g}(\bm{X}_{0}(\phi), \bm{c})\bigr) - \phi \cr \cr
  & = &
  \sum_{k=1}^{M}
  \left. \pderiv{\phi}{X_k} \right|_{\bm{X} = \bm{X}_{0}(\phi)} g_k + 
  \HALF \sum_{k=1}^M \sum_{l=1}^M
  \left. \pnderivM{\phi}{X_k}{X_l} \right|_{\bm{X} = \bm{X}_{0}(\phi)} g_k g_l + \cdots.
\end{eqnarray}
>From the definition of the phase sensitivity function,
  Eq.~(\ref{PhaseSensitivity}), we obtain
\begin{equation}
  \left.\pderiv{\phi}{X_k}\right|_{\bm{X} = \bm{X}_{0}(\phi)} = Z_k(\phi),
\end{equation}
and
\begin{equation}
  \left. \pnderivM{\phi}{X_k}{X_l} \right|_{\bm{X} = \bm{X}_{0}(\phi)}
  = \left. \pderiv{Z_l}{X_k} \right|_{\bm{X} = \bm{X}_{0}(\phi)}
  =  \frac{d Z_l}{d \phi} \left. \frac{\partial \phi}{\partial X_k} \right|_{\bm{X} = \bm{X}_{0}(\phi)}
  = Z_k(\phi) Z_l'(\phi).
\end{equation}
Keeping only terms up to second order in $\bm{g}$, we obtain
\begin{eqnarray}
  \langle G(\phi, \bm{c}) \rangle_{\bm{c}}
  &=& 
  \sum_k Z_k\langle g_k \rangle_{\bm{c}} + \HALF \sum_{k,l} Z_k Z_l'
  \langle g_k g_l \rangle_{\bm{c}},
  \cr \cr
  \langle G(\phi, \bm{c})^2 \rangle_{\bm{c}}
  &=&
  \sum_{k,l} Z_k Z_l \langle g_k g_l \rangle_{\bm{c}}.
\end{eqnarray}
Depending on the picture of the original
Eq.~(\ref{OriginalODE}), the approximate jump magnitude
$\bm{g}(\bm{X}_{0}(\phi), \bm{c})$ for a given mark $\bm{c}$ is
\begin{equation}
  \bm{g}(\bm{X}_{0}(\phi), \bm{c}) =
  \left\{
    \begin{array}{ll}
      \bm{\sigma}(\bm{X}_{0}(\phi), \bm{c}) & \hspace{1mm}\mbox{(Ito)},
      \cr \cr
      \left. \left( e^{\hat{D}} - 1 \right) \bm{X} \right|_{\bm{X} = \bm{X}_{0}(\phi)}& \hspace{1mm}\mbox{(Stratonovich)}.
    \end{array}
  \right.
\end{equation}

For $v(\phi)$ and $D(\phi)$ in the Ito picture of
Eq.~(\ref{OriginalODE}), we obtain
\begin{eqnarray}
  v(\phi) &\cong&
  \omega + \frac{\lambda}{2} \sum_{k,l} Z_k Z_l' \langle \sigma_k \sigma_l \rangle_{\bm{c}},
  \cr \cr
  D(\phi) &\cong& 
  \sum_{k,l} Z_k Z_l \langle \sigma_k \sigma_l \rangle_{\bm{c}},
\end{eqnarray}
where we utilized the assumption $\langle \sigma_k(\phi, \bm{c})
\rangle_{\bm{c}} = \int_{\bm{c}} p(\bm{c}) \sigma_k(\bm{X}_0(\phi),
\bm{c}) d\bm{c} = 0$.
For $v(\phi)$ and $D(\phi)$ corresponding to the Stratonovich picture
of Eq.~(\ref{OriginalODE}), we obtain up to $O(\sigma_k \sigma_l)$,
\begin{eqnarray}
  g_k(\bm{X}_{0}(\phi), \bm{c})
  & \cong &
  \hat{D} X_k|_{\bm{X} = \bm{X}_{0}(\phi)} + \HALF \hat{D}^2 X_k|_{\bm{X} = \bm{X}_{0}(\phi)}
  \cong
  \sigma_k + \HALF \sum_{l} \sigma_l\pderiv{\sigma_k}{X_l},
\end{eqnarray}

\begin{eqnarray}
  g_k(\bm{X}_{0}(\phi), \bm{c}) g_l(\bm{X}_{0}(\phi), \bm{c}) &\cong &\sigma_k\sigma_l,
\end{eqnarray}
so in the Stratonovich picture of Eq.~(\ref{OriginalODE}), we
  obtain
\begin{eqnarray}
  v(\phi) &\cong&
  \omega + \frac{\lambda}{2} \sum_{k,l}
  \left( Z_k Z_l' \langle \sigma_k \sigma_l \rangle_{\bm{c}}
    + Z_k Z_l \langle \sigma_k' \sigma_l \rangle_{\bm{c}}
  \right),
  \cr \cr
  D(\phi) &\cong&
  \sum_{k,l} Z_k Z_l \langle \sigma_k \sigma_l \rangle_{\bm{c}},
\end{eqnarray}
where we used
\begin{equation}
  \sigma_l \pderiv{\sigma_k}{X_l}
  = \sigma_l \pderiv{\sigma_k}{\phi} \frac{\partial \phi}{\partial X_l}
  = \sigma_l Z_l \pderiv{\sigma_k}{\phi}.  
\end{equation}
Now that we have $v(\phi)$ and $D(\phi)$ for both cases, we write an
FPE, and find the corresponding Ito SDE.
The Ito SDE corresponding to the Ito picture, Eq.~(\ref{Ito}),
reads

\begin{eqnarray}
  d\phi(t)
  & = &
  \left( 
    \omega + \frac{\lambda}{2} \sum_{k,l} Z_k Z_l' \langle \sigma_k \sigma_l \rangle_{\bm{c}}
  \right) dt + 
  \sqrt{\lambda} \sum_{k, m} Z_k G_{km} dW_m(t),
\end{eqnarray}
where we introduced $M$ independent Wiener processes $\{ dW_m(t)
  \}$ ($m=1, \cdots, M$)
\footnote{It should be noted that the FPE~(\ref{DiffLimitFPE}) does
  not correspond uniquely to a single SDE, but can correspond to
  several different SDEs with a differing number of noise components,
  which all have the same total magnitude but different number of
  components~\cite{Arnold}.
 Here we introduced $M$ noises, because the original
    Eq.~(\ref{OriginalODE}) is $M$-dimensional, namely, it has $M$
    different directions to be driven by the impulses.
  When discussing the linear stability below, this prescription
  should be used to obtain the correct result.
}
and an $M \times M$ coupling matrix $G_{km}(\phi)$ that
  satisfies
\begin{eqnarray}
  \sum_m G_{km}(\phi_1)  G_{lm}(\phi_2) = \langle \sigma_k(\phi_1) \sigma_l(\phi_2) \rangle_{\bm{c}}.
\end{eqnarray}

Similarly, the Ito SDE corresponding to the Stratonovich
  picture, Eq.(\ref{WhiteNoiseLimit}), leads to
\begin{eqnarray}
  d\phi(t) & = & \left[
    \omega + \frac{\lambda}{2} \sum_{k,l} \left(
      Z_k Z_l' \langle \sigma_k \sigma_l \rangle_{\bm{c}}
      + Z_k Z_l  \langle \sigma_k' \sigma_l \rangle_{\bm{c}}
    \right)
  \right] dt + 
  \sqrt{\lambda} \sum_{k, m} Z_k G_{km} dW_m(t).
\end{eqnarray}
Using the transformation rule between Ito SDE and Stratonovich
  SDE~\cite{Gardiner,Arnold}, this equation can concisely be expressed as a
  Stratonovich SDE
\begin{eqnarray}
  \mbox{(S)} \;\;\;
  d\phi(t) = \omega dt
  + \sqrt{\lambda} \sum_{k, m} Z_k(\phi) G_{km}(\phi) dW_m(t),
\end{eqnarray}
which was the starting point of the previous
  works~\cite{Teramae-Dan,Nakao-Arai-Kawamura}.

\subsection{Linear stability of the synchronized state}

In both pictures of Eq.~(\ref{OriginalODE}), the diffusion-limit
  Ito SDE takes the form
\begin{equation}
  d\phi(t) = \left[ \omega + a(\phi) \right] dt + \sum_m b_m(\phi) dW_m(t),
\end{equation}
 where $a(\phi)$ is periodic in $\phi$, and $b_m(\phi) =
  \sqrt{\lambda} \sum_k Z_k(\phi) G_{km}(\phi)$.
We are interested in the linearized dynamics of the small
  perturbation $\psi(t)$ to $\phi(t)$,
\begin{equation}
  d\psi(t) = \left[ a'(\phi) dt + \sum_m b_m'(\phi) dW_m(t) \right] \psi.
\end{equation}
Using the Ito formula~\cite{Gardiner,Arnold} for changing variables to
$y = \log |\psi|$,
\begin{equation}
  dy(t) = \left[ a'(\phi) - \HALF \sum_m b_m'(\phi)^2 \right] dt + \sum_m b_m'(\phi) dW_m(t).
\end{equation}
The expectation is calculated by replacing the dynamics with the single-oscillator phase PDF, $p(\phi) = 1$,
so the Lyapunov exponent is given as
\begin{equation}
  \Lambda
  =
  - \frac{1}{2} \int_0^1 \sum_m b_m'(\phi)^2 d\phi
  \leq
  0,
  \label{D25}
\end{equation}
where the integral of $a'(\phi)$ vanishes due to the periodicity
  of $a(\phi)$, and the noise term vanishes because
  $b_m'(\phi)$ and $dW_m(t)$ are independent in the Ito SDE and the
expectation of $dW_m(t)$ is $0$.
Therefore, the Lyapunov exponent is the same no matter the
  picture of the SDE Eq.~(\ref{NotClosedPhaseEQ}).
 Inserting $b_m(\phi) = \sqrt{\lambda}\sum_k Z_k(\phi) G_{km}(\phi)$, the
  summation in Eq.~(\ref{D25}) can be calculated as
\begin{eqnarray}
  \sum_m b_m'^2
  &=&
  \lambda \sum_{k, l} Z_k' \left[ \sum_m G_{km} G_{lm} \right] Z_l'
  +
  \lambda \sum_{k, l} Z_k' \left[ \sum_m G_{km} G_{lm}' \right] Z_l \cr \cr \cr
  &&+
  \lambda \sum_{k, l} Z_k \left[ \sum_m G_{km}' G_{lm} \right] Z_l'
  +
  \lambda \sum_{k, l} Z_k \left[ \sum_m G_{km}' G_{lm}' \right] Z_l \cr \cr \cr
  &=&
  \lambda \sum_{k, l}
  \left(
    Z_k' \langle \sigma_k \sigma_l \rangle_{\bm{c}} Z_l'
    +
    Z_k' \langle \sigma_k \sigma_l' \rangle_{\bm{c}} Z_l
    +
    Z_k \langle \sigma_k' \sigma_l \rangle_{\bm{c}} Z_l'
    +
    Z_k \langle \sigma_k' \sigma_l' \rangle_{\bm{c}} Z_l
  \right),
\end{eqnarray}
where we used
\begin{eqnarray}
  \langle \sigma_k' \sigma_l \rangle_{\bm{c}}(\phi)
  = \left. \frac{\partial}{\partial \phi_1} \langle \sigma_k(\phi_1)
    \sigma_l(\phi_2) \rangle_{\bm{c}} \right|_{(\phi_1, \phi_2) = (\phi, \phi)},
\end{eqnarray}
etc., so we finally obtain
\begin{eqnarray}
  \Lambda
  = - \frac{\lambda}{2}
  \sum_{k, l}
  \int_0^1 \left[
    Z_k' Z_l' \langle \sigma_k \sigma_l \rangle_{\bm{c}} +
    2 Z_k' Z_l \langle \sigma_k \sigma_l' \rangle_{\bm{c}} +
    Z_k Z_l \langle \sigma_k' \sigma_l' \rangle_{\bm{c}}
  \right].
\end{eqnarray}
This expression coincides with the approximate Lyapunov exponent that
we obtained by a Taylor expansion in
Eq.~(\ref{LyapunovFromG_TaylorExp}), and gives a multiplicative
generalization to the previous results obtained by Teramae and Tanaka
in Ref.~\cite{Teramae-Dan} (our result in
Ref.~\cite{Nakao-Arai-Kawamura} includes this result).



\section{Stuart-Landau oscillator}

\begin{figure}[htbp]
  \centering
  \includegraphics*[width=0.8\hsize]{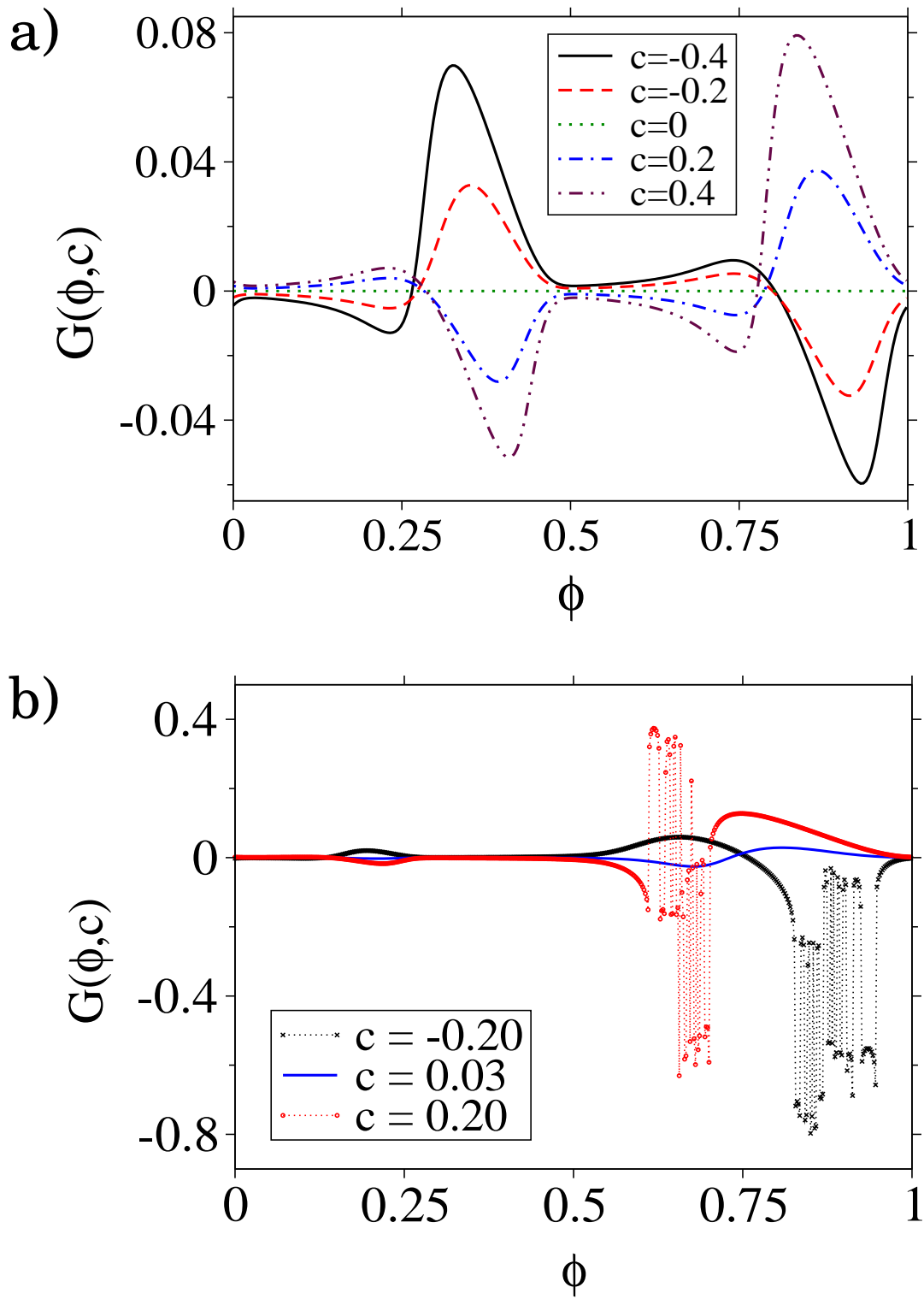}
  \caption{(Color online) (a) PRCs for FHN with $I_0 = 0.8$ and additive impulse
    intensities $c \in [-0.4, 0.4]$, (b)PRCs for FHN with $I_0 = 0.34$
    and additive impulse intensities $c = -0.20, 0.03, 0.20$.}
  \label{Fig:FHNPRC}
\end{figure}

In Sec. III, we discussed the case in which the response of the
oscillator to sufficiently strong perturbations result in PRCs that
appear jagged, Fig.~\ref{Fig:FHNPRC}b).  For such PRCs, the
desynchronization is intuitive: if a nearly-synchronized group of
oscillators near such a jagged response receive an common impulse,
they end up with widely distributed phases~\cite{Nagai-Nakao,
  Nakao-Arai}.
In Ref.~\cite{Goldobin-Pikovsky}, the same situation is described
differently, where the importance of the "heavy-tails" of the
distribution of relaxation rates of transverse perturbations for
oscillators near the bifurcation point is emphasized.

However, the PRC need not have such a pathologic shape for
desynchronization.  It may even be sinusoidal as shown in
Ref.~\cite{Pikovsky-R-K1}, Sec. 15.
Such a case occurs with the Stuart-Landau (SL) oscillator, which
describes the small-amplitude oscillations near the supercritical Hopf
bifurcation point of a general system of ODEs~\cite{Kuramoto}.

\begin{figure}[htbp]
  \centering
  \includegraphics*[width=0.8\hsize]{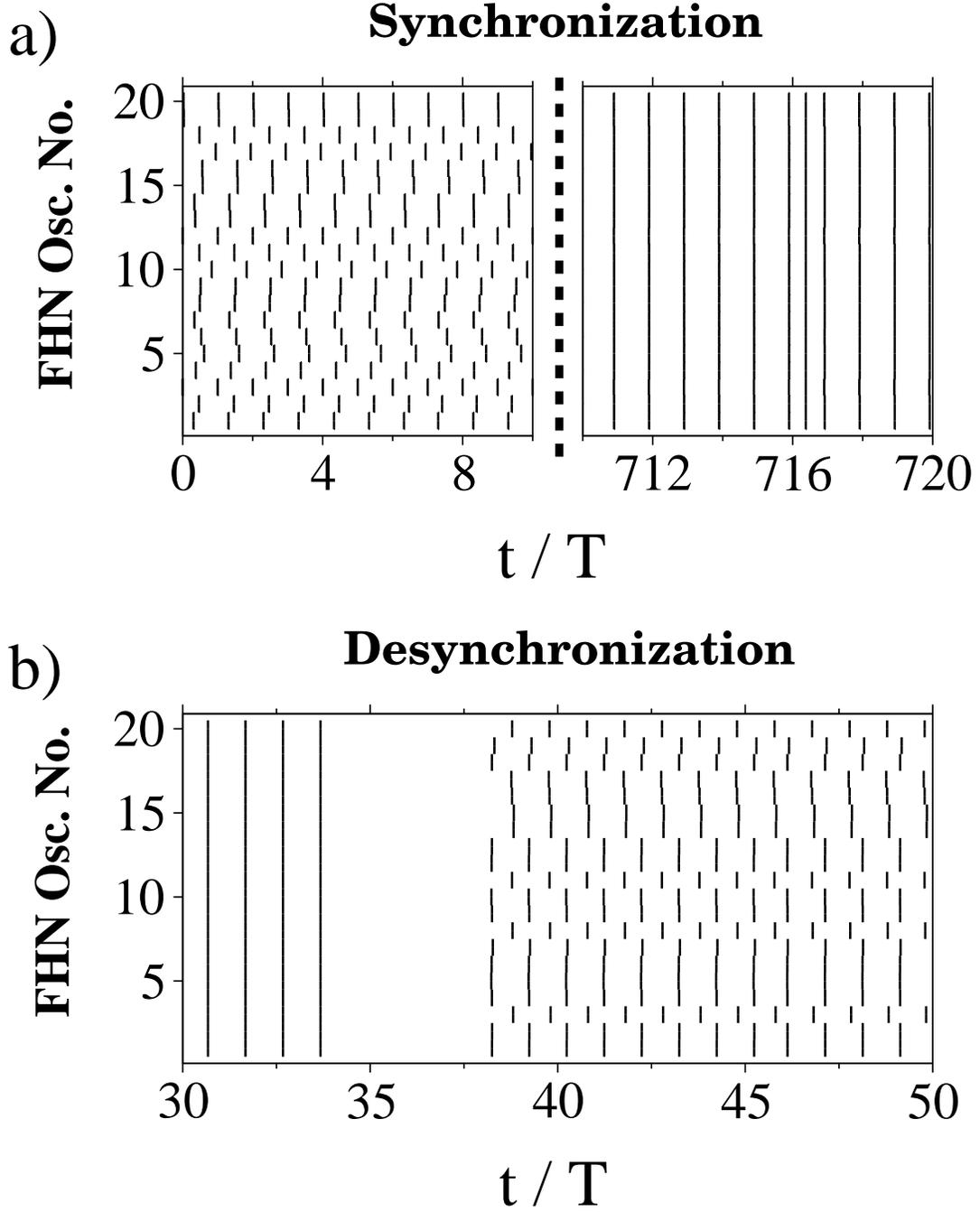}
  \caption{Raster plots of FHN oscillators showing synchronization and
    desynchronization due to common impulses with rate $\lambda =
    1/4T$ for both.  Time axis normalized by natural frequency of
    oscillators.  Each tick indicates the time oscillator passed
    through $\phi = 0$ near the limit cycle.  (a) Synchronization of
    FHN ($I_0 = 0.8$, $c = 0.3$, $D = 5 \times 10^{-6}$) and (b)
    Desynchronization of FHN ($I_0 = 0.34$, $c = 0.2$, $D = 5 \times
    10^{-8}$).  The wide blank without ticks in (b) corresponds to the
    situation where the orbit is transiently trapped around the
    unstable focus.}
  \label{Fig:FHNRasterPlots}
\end{figure}

\begin{figure}[htbp]
  \centering
  \includegraphics*[width=0.8\hsize]{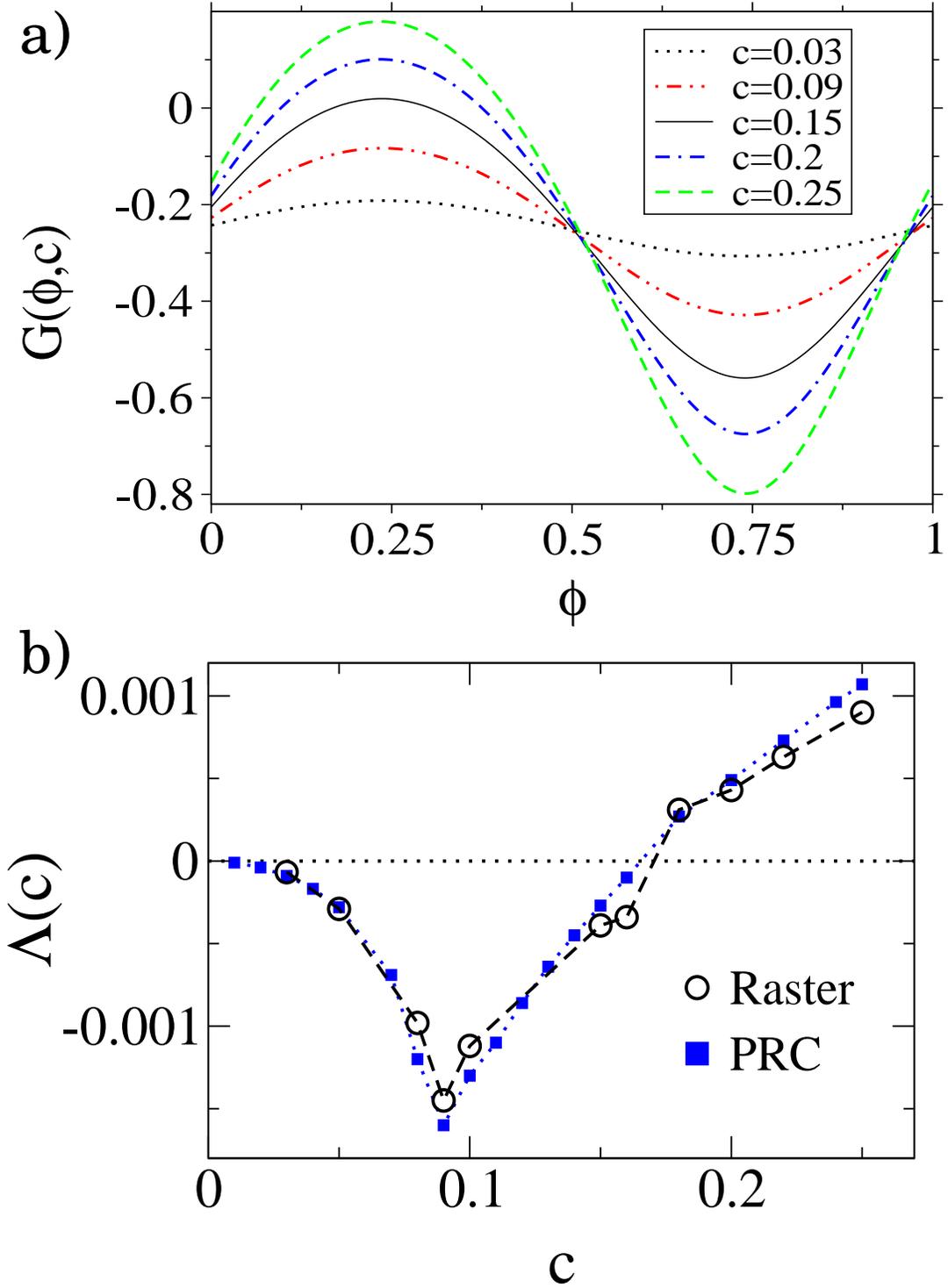}
  \caption{(Color online)  a) PRCs for SL model ($c_0 = -c_2 = 12$) for various
      additive impulse intensities.  b) Comparison of the Lyapunov
      exponents $\Lambda$ between the direct measurement from the
      raster plot and the theoretical prediction from the PRC for SL
      oscillators driven by additive impulses of intensity $c$, and
      Poisson impulse rate $\lambda = 1/380T$.  We chose a large
      inter-impulse interval (380T) because the SL oscillators have a
      very slow relaxation back to the limit-cycle orbit following a
      perturbation.}
  \label{Fig:SLFigs}
\end{figure}

Consider the following SL oscillator driven by random Poisson
impulses:
\begin{eqnarray}
  \dot{u} & = & (u - c_0 v) - (u - c_2 v)(u^2 + v^2)
  + \sigma(v, c) \sum_{n=1}^{N(t)} h(t - t_{n})\nonumber \\
  \dot{v} & = & (v + c_0 u) - (v + c_2 u)(u^2 + v^2),
\end{eqnarray}
where $h(t)$ and $\sigma(v, c)$ as described above for the FHN
oscillator.
For comparison with the FHN oscillator, we followed the same procedure
for the SL oscillators and found the existence of synchronizing and
desynchronizing impulse strengths (raster data not shown but are
qualitatively similar to Fig.~\ref{Fig:FHNRasterPlots}).
PRCs are shown in Fig.~\ref{Fig:SLFigs}(a) and Lyapunov exponents
obtained using Eq.~(\ref{LyapunovFromG}) and by direct measurement are
shown in Fig.~\ref{Fig:SLFigs}(b).
The PRCs are almost sinusoidal but slightly deformed because the
impulse intensity $c$ is finite.
As expected, the Lyapunov exponent $\Lambda(c)$ shown in
Fig.~\ref{Fig:SLFigs}(b) is qualitatively very similar to that
obtained in Ref.~\cite{Pikovsky-R-K1} calculated for a circle map with
a sinusoidal PRC receiving common Poisson impulses, which predicts
synchronization for weak impulses and desynchronization for stronger
impulses.

Note that in our present treatment of Poisson-driven limit cycles, we
do not need to discuss the FHN-type oscillator and the SL-type
oscillator separately.  We can simply adopt the same one-dimensional
phase model with the standard definition of the PRC, which
quantitatively predicts the Lyapunov exponent in both synchronization
and desynchronization regimes.

\clearpage

\end{document}